\documentclass[10pt,journal,compsoc]{IEEEtran}
\ifCLASSOPTIONcompsoc
  \usepackage[nocompress]{cite}
\else
  \usepackage{cite}
\fi
\usepackage{dblfloatfix}
\usepackage{graphicx}
\usepackage{epstopdf}
\usepackage{subfigure}
\usepackage{multirow}
\usepackage{algorithmic}
\usepackage{algorithm}
\usepackage{amsmath}
\usepackage{color}
\usepackage{float}
\usepackage{CJK}
\usepackage{subfigure}
\usepackage{threeparttable}
\usepackage[none]{hyphenat}
\ifCLASSINFOpdf
\else
\fi
\begin{document}
%
\title{A Survey on Incorporating Domain Knowledge into Deep Learning for Medical Image Analysis}
%
%
%
%

\author{Xiaozheng~Xie,
        Jianwei~Niu,~\IEEEmembership{Senior~Member,~IEEE},
        Xuefeng~Liu,
        Zhengsu~Chen,
        Shaojie~Tang,~\IEEEmembership{Member,~IEEE}
        and Shui Yu
\IEEEcompsocitemizethanks{\IEEEcompsocthanksitem X. Xie, J. Niu, X. Liu and Z. Chen are with the State Key Laboratory of Virtual Reality Technology and Systems, School of Computer Science and Engineering, Beihang University, Beijing 100191, China.
E-mails: \{xiexzheng,niujianwei,liu\_xuefeng,danczs\}@buaa.edu.cn
\IEEEcompsocthanksitem J. Niu is also with the Beijing Advanced Innovation Center for Big Data and Brain Computing (BDBC) and Hangzhou Innovation Institute of Beihang University.
\IEEEcompsocthanksitem S. Tang is in Jindal School of Management, The University of Texas at Dallas. E-mails: tangshaojie@gmail.com
\IEEEcompsocthanksitem S. Yu is in School of Computer Science, University of Technology Sydney, Australia. E-mails: Shui.Yu@uts.edu.au
}

}

\IEEEtitleabstractindextext{%
\begin{abstract}
Although deep learning models like CNNs have achieved great success in medical image analysis, the small size of medical datasets remains a major bottleneck in this area. To address this problem, researchers have started looking for external information beyond current available medical datasets. Traditional approaches generally leverage the information from natural images via transfer learning. More recent works utilize the domain knowledge from medical doctors,
to create networks that resemble how medical doctors are trained,
mimic their diagnostic patterns, or focus on the features or areas they pay particular attention to. In this survey, we summarize the current progress on integrating medical domain knowledge into deep learning models for various tasks, such as disease diagnosis, lesion, organ and abnormality detection, lesion and organ segmentation. For each task, we systematically categorize different kinds of medical domain knowledge that have been utilized and their corresponding integrating methods. We also provide current challenges and directions for future research.
\end{abstract}

}

\maketitle
\IEEEdisplaynontitleabstractindextext
%
\IEEEpeerreviewmaketitle

\IEEEraisesectionheading{\section{Introduction}
\label{sec:introduction}}
%
%
%
%
\begin{CJK}{UTF8}{gbsn}

\IEEEPARstart{R}{ecent} years have witnessed a tremendous progress in computer-aided detection/diagnosis (CAD) in medical imaging and diagnostic radiology, primarily thanks to the advancement of deep learning techniques. Having achieved great success in computer vision tasks, various deep learning models, mainly convolutional neural networks (CNNs), soon be applied to CAD. Among the applications are the early detection and diagnosis of breast cancer, lung cancer, glaucoma, and skin cancer \cite{esteva2017dermatologist,8471199,zhou2017fine-tuning,zhu2018deeplung}.

However, the small size of medical datasets continues to be an issue in obtaining 
satisfactory deep learning model for CAD; in general,  
bigger datasets result in better deep learning models \cite{halevy2009unreasonable}. In traditional computer vision tasks, there are many large-scale and well-annotated datasets, such as ImageNet
\footnote{http://www.image-net.org/}
(over 14M labeled images from 20k categories) and COCO
\footnote{http://mscoco.org/}
(with more than 200k annotated images across 80 categories).  In contrast, some popular publicly available medical datasets are much smaller (see Table \ref{tab:medical_dataset}). For example, among the datasets for different tasks, only ChestX-ray14 and DeepLesion, contain more than 100k labeled medical images, while most datasets only have a few thousands or even hundreds of medical images.

\begin{table*}[]
\center
\small
\caption{Examples of popular datasets in the medical domain}
\begin{tabular}{|l|l|l|l|l|}
\hline
Name        & Purpose      & Type & Imaging    & Number of Images                                                                                            \\ \hline
\begin{tabular}[c]{@{}c@{}}ADNI \cite{WEINER2017561}\end{tabular}        &Classification          & Brain   &Multiple       & 1921 patients                                                                                      \\ \hline
\multirow{1}{*}{\begin{tabular}[c]{@{}c@{}}ABIDE \cite{di2014autism}\end{tabular}}        & \multirow{1}{*}{Classification}     & \multirow{1}{*}{Brain}  & \multirow{1}{*}{MRI}          & 539 patients and 573 controls \\  \hline
\begin{tabular}[c]{@{}c@{}}ACDC \cite{acdc_web}\end{tabular}        &Classification     & Cardiac    &MRI            & 150 patients                                                                                      \\ \hline
\multirow{1}{*}{\begin{tabular}[c]{@{}c@{}}ChestX-ray14 \cite{article}\end{tabular}}     &\multirow{1}{*}{Detection} & \multirow{1}{*}{Chest}   &\multirow{1}{*}{X-ray}          & 112,120 images from 30,805 patients \\  \hline
\begin{tabular}[c]{@{}c@{}}LIDC-IDRI \cite{armato2011lung}\end{tabular}    &Detection      & Lung  &CT, X-ray      & 1,018 patients                                                                                     \\  \hline
\begin{tabular}[c]{@{}c@{}}LUNA16 \cite{luna2016}\end{tabular}    &Detection     & Lung   &CT             & 888 images   \\  \hline
\multirow{1}{*}{\begin{tabular}[c]{@{}c@{}} MURA \cite{rajpurkar2017mura}\end{tabular}}    &\multirow{1}{*}{Detection}   & \multirow{1}{*}{\begin{tabular}[c]{@{}c@{}}Musculo-skeletal\end{tabular}}  &\multirow{1}{*}{X-ray}          & \begin{tabular}[c]{@{}c@{}}40,895 images from 14,982 patients\end{tabular}                                                                 \\  \hline
\begin{tabular}[c]{@{}c@{}}BraTS2018 \cite{bakas2018identifying}\end{tabular}      &Segmentation    & Brain    &MRI            & 542 images 
\\  \hline
\begin{tabular}[c]{@{}c@{}}STARE \cite{hoover2000locating}\end{tabular}     &Segmentation  & Eye  &SLO            & 400 images                                                                                        \\ \hline
\begin{tabular}[c]{@{}c@{}}DDSM \cite{ddsm}\end{tabular}         &\begin{tabular}[c]{@{}c@{}}Classification\\Detection\end{tabular}       & Breast   & Mammography    & 2,500 patients
\\ \hline
\begin{tabular}[c]{@{}c@{}}DeepLesion \cite{yan2018deeplesion}\end{tabular}   &\begin{tabular}[c]{@{}c@{}}Classification\\Detection\end{tabular}      & Multiple &CT             & 32,735 images from 4,427 patients                                                              \\  \hline
\begin{tabular}[c]{@{}c@{}}Cardiac MRI \cite{cardiacMRI}\end{tabular}  &\begin{tabular}[c]{@{}c@{}}Classification\\Segmentation\end{tabular}      & Cardiac    &MRI           & 7,980 images from 33 cases                                                                         \\ \hline
\begin{tabular}[c]{@{}c@{}}ISIC 2018 \cite{codella2019skin}\end{tabular}       &\begin{tabular}[c]{@{}c@{}}Classification\\Detection\\Segmentation \end{tabular}      & Skin   & Dermoscopic    & 13,000 images                                                                                      \\                                                              \hline
\end{tabular}
\label{tab:medical_dataset}
\end{table*}

The lack of medical datasets is represented in three aspects. First, the number of medical images in datasets is usually small. This problem is mainly due to the high cost associated with the data collection. Medical images are collected from computerized tomography (CT), Ultrasonic imaging (US), magnetic resonance imaging (MRI) scans, positron emission tomography (PET), all of which are expensive and labor-intensive. Second, only a small portion of medical images are annotated. These  annotations including classification labels (e.g., benign or malignant), the segmentation annotations of lesion areas, etc., require efforts from experienced doctors. Third, it is difficult to collect enough positive cases for some rare diseases to obtain the balanced datasets.

One direct consequence of the lack of well annotated medical data is that the trained deep learning models can easily suffer from  the overfitting problem \cite{chen2016deep}. As a result, the models perform very well on training datasets, but fail when dealing with new data from the problem domain. Correspondingly, many existing studies on medical image analysis adopt techniques from computer vision to address overfitting, such as reducing the complexity of the network \cite{ciompi2015automatic,shin2016deep}, adopting some regularization techniques \cite{erickson2017machine}, or using data augmentation strategies \cite{zhu2017unpaired}.

However, in essence, both decreasing model complexity and leveraging data augmentation techniques only focus on the target task on the given datasets, but \emph{do not introduce any new information into deep learning models}. Nowdays,  introducing more information, beyond the given medical datasets has become a more promising approach to address the problem of small-sized medical datasets.  

The idea of introducing external information to improve the performance of deep learning models for CAD is not new. For example, it is common practice to first train a deep learning model on some natural image datasets like ImageNet, and then  fine tune them on target medical datasets \cite{huynh2016mo}. This process, called transfer learning \cite{pan2009survey}, implicitly introduces information from natural images. Besides natural images, multi-modal medical datasets or medical images from different but related diseases can also be used to improve the performance of deep learning models \cite{samala2018breast,yu2019annotation}.


Moreover, as experienced medical doctors (e.g., radiologists, ophthalmologists, and dermatologists) can generally give fairly accurate results, it is not surprising that their knowledge may help  deep learning models to better accomplish the designated tasks. The domain knowledge of medical doctors includes the way they browse images, the particular areas they usually focus on, the features they give special attention to, and the anatomical prior knowledge they used. These types of knowledge are accumulated, summarized, and validated by a large number of practitioners over many years based on a huge amount of cases. 
Note that in this survey any network that incorporate one of these types of  knowledge in their training or designing process should be regarded as the one incorporated medical domain knowledge.

\begin{figure*}[htp!] \begin{centering}
   \includegraphics[width=0.9\linewidth]{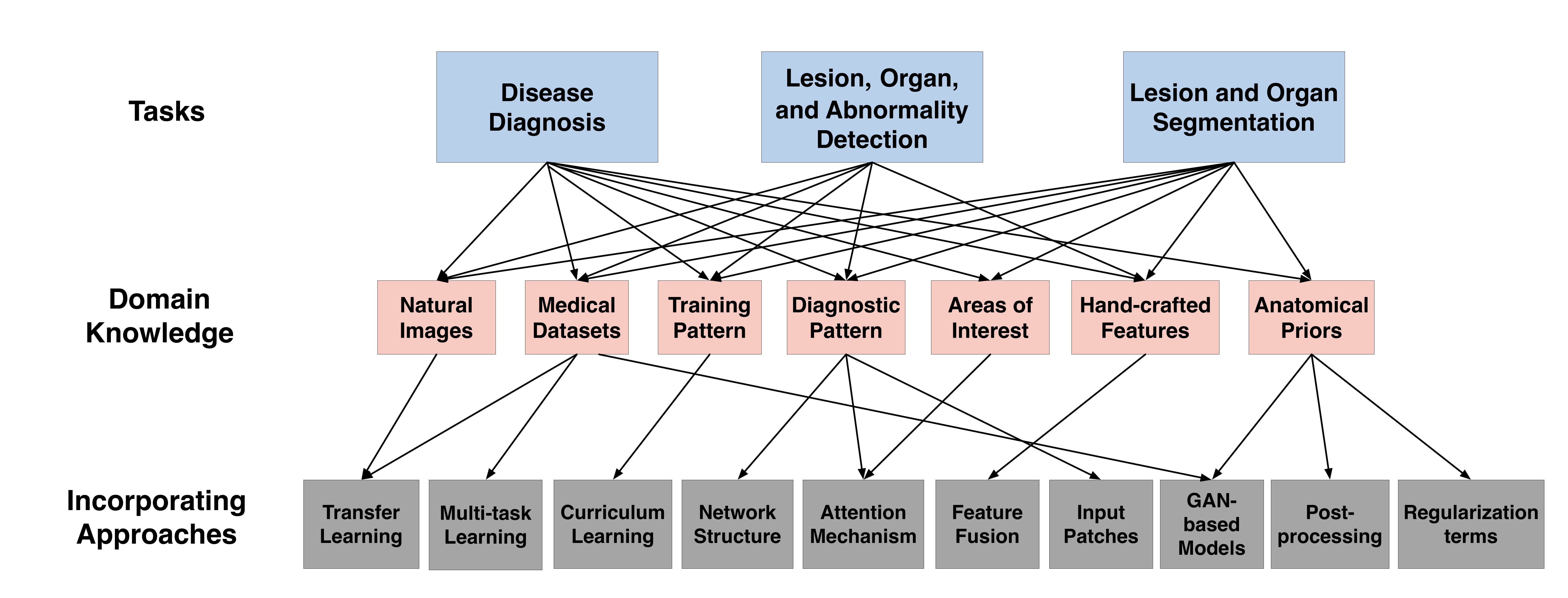}
   \centering
   \caption{Methods of information categorization and incorporating methods in disease diagnosis; lesion, organ, and abnormality detection; lesion and organ segmentation.
   }
   \label{fig:1}
   \end{centering}
 \end{figure*}
In this survey, we focus on the three main tasks of medical image analysis: (1) disease diagnosis, (2) lesion, organ and abnormality detection, and (3) lesion and organ segmentation. We also include other related tasks such as the image reconstruction, image retrieval and report generation. This survey demonstrates that, for almost all tasks, identifying and carefully integrating one or more types of domain knowledge related to the designated task will improve the performance of deep learning models.
We organize existing works according to the following three aspects: the types of tasks, the types of domain knowledge that are introduced, and the ways of introducing the domain knowledge.



More specifically, in terms of the types of domain knowledge, 
some of them are of high-level such as training pattern \cite{maicas2018training,tang2018attention,wang2018deep} and diagnostic pattern. Some domain knowledge are low-level, such as particular features and special areas where medical doctors pay more attention to \cite{hsu2019breast}. In particular, in disease diagnosis tasks, high-level domain knowledge is widely utilized. For an object detection task, the low-level domain knowledge, such as detection patterns and specific features where medical doctors give special attention is more commonly adopted. For lesion or organ segmentation tasks, anatomical priors and the knowledge from different modalities seem to be more useful \cite{chen2019learning,larrazabal2019anatomical,valverde2019one-shot}.

\begin{figure*}[hbp!] \begin{centering}
   \includegraphics[width=0.9\linewidth]{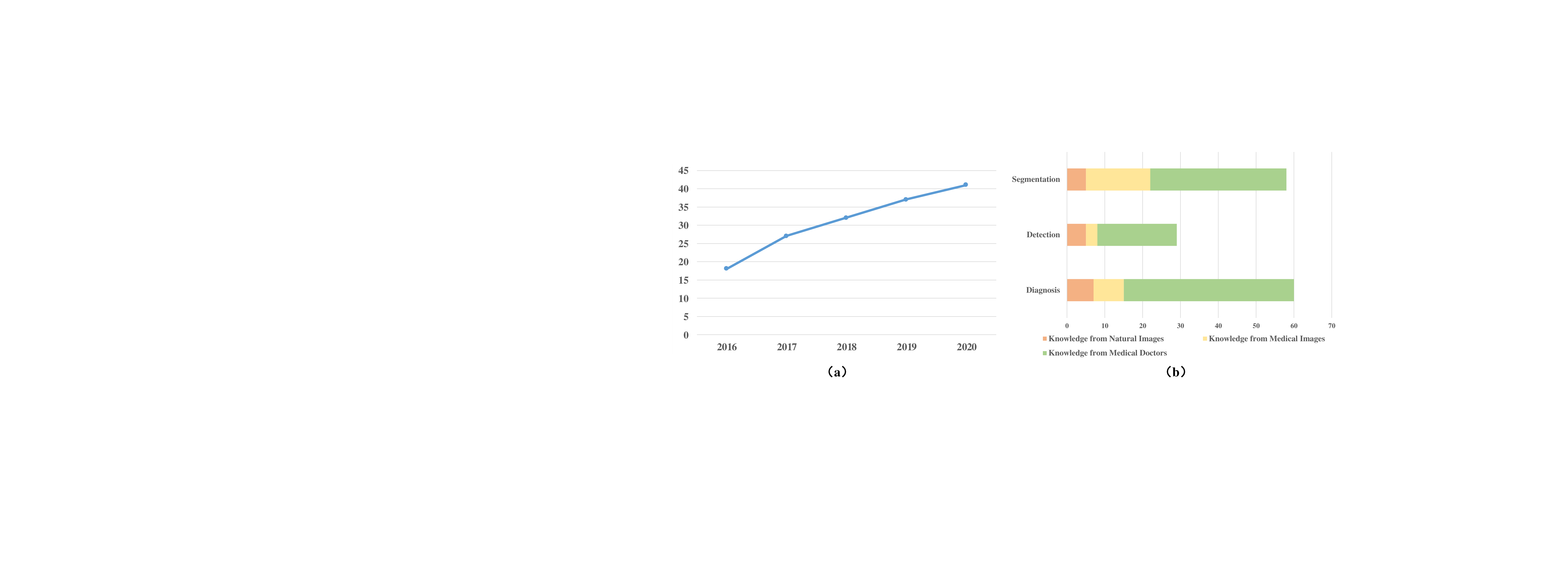}
   \centering
   \caption{(a) Number of papers arranged chronically (2016-2020). (b) The distribution of selected papers in different applications of medical image analysis.
   }
   \label{fig:statis}
   \end{centering}
 \end{figure*}

In terms of the integrating methods, various approaches have been designed to incorporate different types of domain knowledge into networks \cite{ji2020infant}. For example, a simple approach is to concatenate  hand-crafted features with the ones extracted from deep learning models \cite{xie2016lung}. In some works, network architectures are revised to simulate the pattern of radiologists when they read images \cite{Diagnose}. Attention mechanism, which allows a network to pay more attention to a certain region of an image, is a powerful technique to incorporate domain knowledge of radiologists \cite{li2019attention}. In addition, multi-task learning and meta learning are also widely used to introduce medical domain knowledge into deep learning models \cite{chen2016automatic,yue2019cardiac}.

Although there are a number of reviews on deep learning for medical image analysis, including \cite{debelee2019survey,litjens2017survey,shen2017deep,suzuki2017survey}, they all describe the existing works from the application point of view, i.e., how deep learning techniques are applied to various medical applications. To the best of our knowledge, there is no review that gives systematic introduction on \emph{how medical domain knowledge can help deep learning models}. This aspect, we believe, is the unique feature that distinguishes deep learning models for CAD from those for general computer vision tasks.

Fig. \ref{fig:1} gives the overview on how we organize the related researches. At the top level, 
existing studies are classified into three main categories according to their purposes: (1) disease diagnosis, (2) lesion, organ and abnormality detection, and (3) lesion and organ segmentation.
In each category, we organize them into several groups based on the types of extra knowledge have been incorporated. At the bottom level, they are further categorized according to the different integrating approaches of the domain knowledge.

This survey contains more than 200 papers (163 are with domain knowledge), most of which are published recently (2016-2020), on a wide variety of applications of deep learning techniques for medical image analysis. 
In addition, most of the corresponding works are from the conference proceedings for MICCAI, EMBC, ISBI and some journals such as TMI, Medical Image Analysis, JBHI and so on. 
\begin{figure*}[htp!] \begin{centering}
   \includegraphics[width=0.65\linewidth]{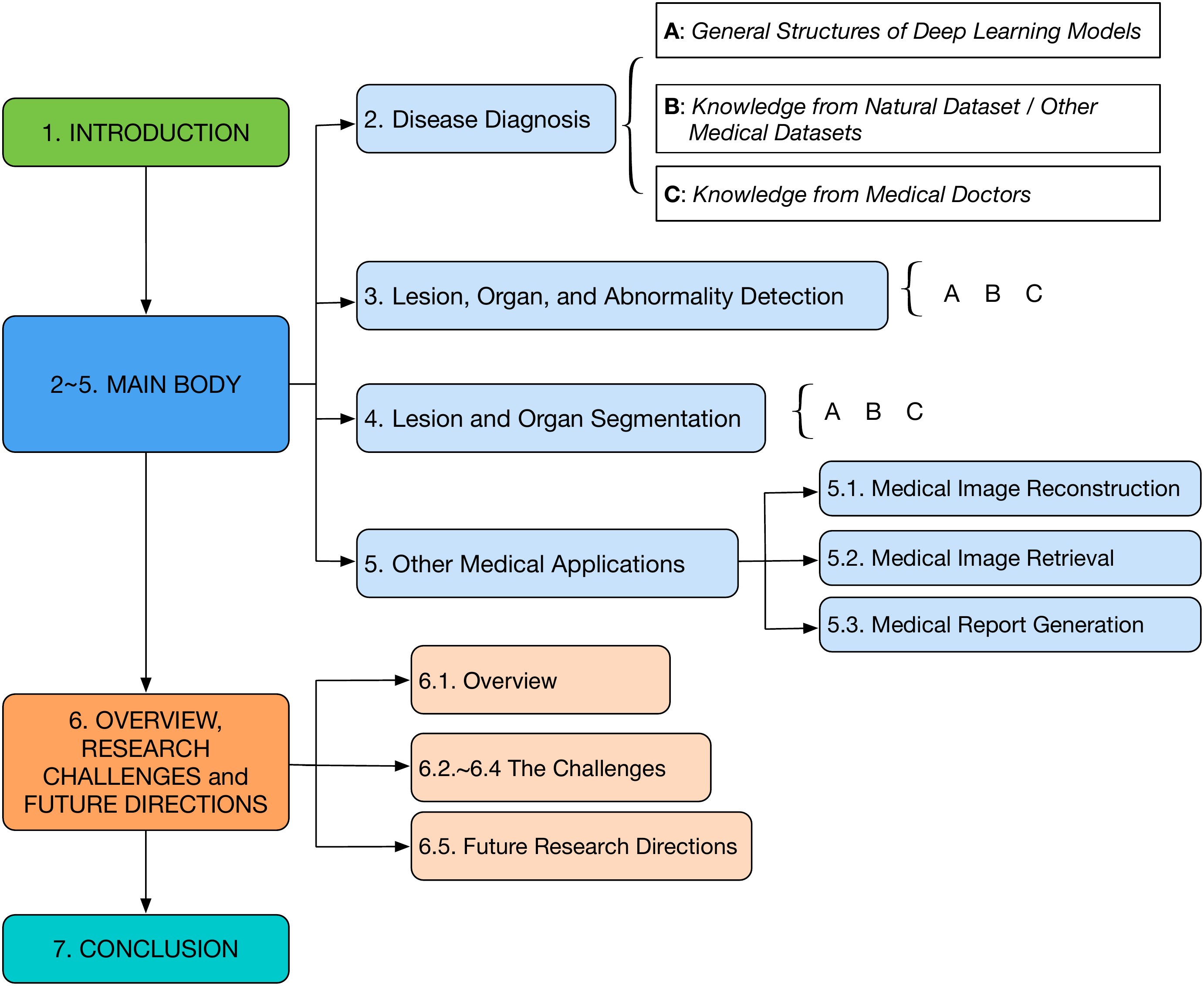}
   \centering
   \caption{The organizational structure of this survey.
   }
   \label{fig:structure}
   \end{centering}
 \end{figure*}

The distribution of these papers are shown in Fig. \ref{fig:statis}(a). It can be seen that the number of papers increases rapidly from 2016 to 2020. With respect to the applications, most of them are related to disease diagnosis and lesion/organ segmentation (shown in Fig. \ref{fig:statis}(b)). To sum up, with this survey we aim to:
\begin{itemize}
\item summarize and classify different types of domain knowledge in medical areas that are utilized to improve the performance of deep learning models in various applications;
\item summarize and classify different ways of introducing medical domain knowledge into deep learning models;
\item give the outlook of challenges and future directions in integrating medical domain knowledge into deep learning models.
\end{itemize}

The rest of the survey is organized as follows. Sections \ref{sec:diagnosis}, \ref{sec:detection} and \ref{sec:segmentation} introduce the existing works for the major three tasks in medical image analysis. Besides these three major tasks, other tasks in medical image analysis are described in Section \ref{sec:otherappl}. In each section, we first introduce the general architectures of deep learning models for a task, and then categorize related works according to the types of the domain knowledge to be integrated. Various incorporating methods for each type of domain knowledge are then described. Section \ref{sec:future_work} discusses research challenges, and gives the outlook of future directions. Lastly, Section \ref{sec:conclusion} concludes this survey. The structure of this survey is shown in Fig. \ref{fig:structure}.



\section{Disease Diagnosis}
\label{sec:diagnosis}

Disease diagnosis refers to the task of determining the type and condition of possible diseases based on the medical images provided. In this section, we give an overview of the deep learning models that generally used for disease diagnosis. Concretely, subsection \ref{sec:diagmodel} outlines the general structures of deep learning models used for disease diagnosis. Subsection \ref{sec:diag_natural} introduces the works that utilize knowledge from natural images or other medical datasets. Deep learning models that leverage knowledge from medical doctors are introduced in Subsection \ref{sec:diag_rediologists} in detail. Lastly, Subsection \ref{sec:diag_overview} summarizes the research of disease diagnosis.

\subsection{General Structures of Deep Learning Models Used for Disease Diagnosis}
\label{sec:diagmodel}

In the last decades, deep learning techniques, especially CNNs, have achieved a great success in disease diagnosis.

\begin{figure}[htp!] \begin{centering}
   \includegraphics[width=1\linewidth]{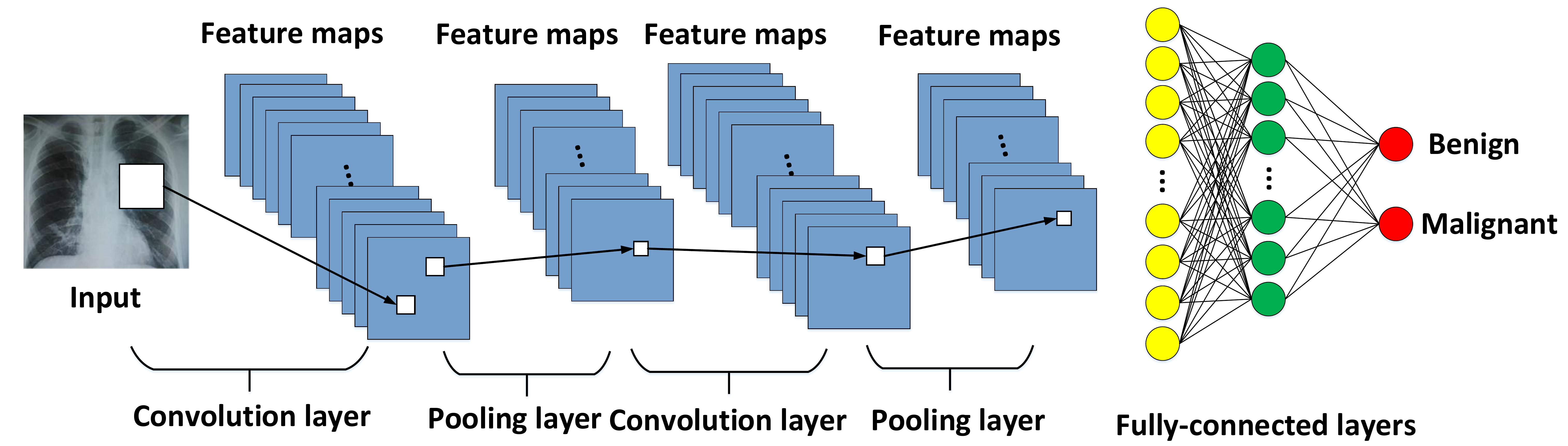}
   \centering
   \caption{A typical CNN architecture for medical disease diagnosis.
   }
   \label{fig:cnn}
   \end{centering}
 \end{figure}

Fig. \ref{fig:cnn} shows the structure of a typical CNN that used for disease diagnosis in chest X-ray image. The CNN employs alternating convolutional and pooling layers, and contains trainable filter banks per layer. Each individual filter in a filter bank is able to generate a feature map. This process is alternated and the CNN can learn increasingly more and more abstract features that will later be used by the fully connected layers to accomplish the classification task.

Different types of CNN architectures, from AlexNet \cite{krizhevsky2012imagenet}, GoogLeNet \cite{szegedy2015going}, VGGNet \cite{simonyan2015very}, ResNet \cite{he2016deep} to DenseNet \cite{huang2017densely}, have achieved a great success in the diagnosis of various diseases. For example,
GoogLeNet, ResNet, and VGGNet models are used in the diagnosis of canine ulcerative keratitis \cite{kim2019cnn}, and most of them achieve accuracies of over 90\% when classifying superficial and deep corneal ulcers. DenseNet is adopted to diagnose lung nodules on chest X-ray radiograph \cite{li2019multi}, and experimental results show that more than 99\% of lung nodules can be detected. In addition, it is found that VGGNet and ResNet are more effective than other network structures for many medical diagnostic tasks \cite{Diagnose,liu2018integrate,mitsuhara2019embedding,xie2019knowledge-based}.


However, the above works generally directly apply CNNs to medical image analysis or slightly modified CNNs (e.g., by modifying the number of kernals, the number of channels or the size of filters), and no medical knowledge is incorporated. In addition, these methods generally require large medical datasets to achieve a satisfactory performance.



In the following subsections, we systematically review on the research that utilizes  medical domain knowledge for the disease diagnosis. The types of knowledge and the incorporating methods are summarized in Table \ref{tab:diagnosis_knowledge}.
\begin{table*}[]
\small
\center
\caption{A compilation of the knowledge and corresponding incorporating methods used in disease diagnosis}
\begin{tabular}{|c|l|l|l|}
\hline
\begin{tabular}[c]{@{}l@{}}Knowledge Source\end{tabular}                 & \begin{tabular}[c]{@{}l@{}}Knowledge Type\end{tabular}                                                                                                             & \begin{tabular}[c]{@{}l@{}}Incorporating Method\end{tabular}                                                                                                                                                & References                                                                                                                                                                                                                                                                                                                                                                                                                                                   \\ \hline
\begin{tabular}[c]{@{}l@{}}Natural datasets\end{tabular}                   & \begin{tabular}[c]{@{}l@{}}Natural images\end{tabular}                                                                                                            & \begin{tabular}[c]{@{}l@{}}Transfer learning\end{tabular}                                                                                                                                                   & 
{\begin{tabular}[c]{@{}l@{}}\cite{bar2015deep}\cite{esteva2017dermatologist} \cite{wimmer2016convolutional}\cite{cao2018improve}\cite{wang2017chestx} \cite{hussein2017risk} \cite{li2018path}\cite{huynh2016digital} \end{tabular}} \\ \hline
\multirow{2}{*}{\begin{tabular}[c]{@{}l@{}}Medical datasets\end{tabular}} & \begin{tabular}[c]{@{}l@{}}Multi-modal images\end{tabular}                                                                                                         & \begin{tabular}[c]{@{}l@{}}Transfer learning\end{tabular}                                                                                     & \begin{tabular}[c]{@{}l@{}}\cite{samala2018breast}\cite{hadad2017classification} \cite{samala2017multi}\cite{azizi2017transfer}\cite{li2020digital}\cite{han2020deep}\end{tabular}
                                                                                                                                                                                                                                                                                                                                                                                                                                                                                                                                                    \\ \cline{2-4}
                                 & \begin{tabular}[c]{@{}l@{}}Datasets from other diseases\end{tabular}                                                                                               & \begin{tabular}[c]{@{}l@{}}Multi-task learning\end{tabular}                                                                                                                                                 & \begin{tabular}[c]{@{}l@{}}\cite{li2019canet}\cite{liao2019multi} \end{tabular}                                                                                                                                                                                                                                                                                                                                                                                                                                                           \\ \hline
\multirow{9}{*}{\begin{tabular}[c]{@{}l@{}}Medical doctors\end{tabular}} & \begin{tabular}[c]{@{}l@{}}Training pattern\end{tabular}                                                                                                           & \begin{tabular}[c]{@{}l@{}}Curriculum learning\end{tabular}                                                                                                                                                 &\begin{tabular}[c]{@{}l@{}}\cite{maicas2018training}\cite{tang2018attention} \cite{jimenez2019medical}\cite{haarburger2019multi}\cite{zhao2020egdcl}
\cite{jimenez2020curriculum}\cite{wei2020learn}\cite{qi2020curriculum}\end{tabular}                                                                                                                                                                                                                                                                                                                                                                                                                                                              \\ \cline{2-4}
                                 & \begin{tabular}[c]{@{}l@{}}Diagnostic patterns\end{tabular}                                                                                                         & \begin{tabular}[c]{@{}l@{}}Network  design\end{tabular}                                                                                                                                         & \begin{tabular}[c]{@{}l@{}}\cite{Diagnose}\cite{gonzalez2018dermaknet}
                                 \cite{wang2020learning}\cite{huang2020dual}
                                 \cite{yang2020momminet}\cite{liu2020semi}\end{tabular}                                                                                                                                                                                                                                                                                                                                                                                                                                                             \\ \cline{2-4}
                                 & \begin{tabular}[c]{@{}l@{}}Areas doctors focus on\end{tabular}                                                                                                    & \begin{tabular}[c]{@{}l@{}}Attention  mechanism 
                                 \end{tabular}                                                                & \begin{tabular}[c]{@{}l@{}}\cite{li2019attention} \cite{mitsuhara2019embedding}  \cite{8637959}\cite{cui2020collaborative}
                                 \cite{xiedg}\cite{zhang2020short}\end{tabular}                                                                                                                                                                                                                                                                                                                                                                                                                                                             \\ \cline{2-4}
                                 & \multirow{4}{*}{\begin{tabular}[c]{@{}l@{}}Features doctors  focus on\end{tabular}}                                       & \begin{tabular}[c]{@{}l@{}}Decision level fusion\end{tabular}  &  \begin{tabular}[c]{@{}l@{}}\cite{huynh2016digital} \cite{moradi2016hybrid} \cite{majtner2016combining}\cite{xie2018fusing} \cite{antropova2017deep}\cite{xia2020comparison} \end{tabular}                               \\ \cline{3-4}
                                   & &\begin{tabular}[c]{@{}l@{}}Feature level fusion\end{tabular} & 
                                   {\begin{tabular}[c]{@{}l@{}}\cite{xie2016lung}\cite{hagerty2019deep} \cite{cao2018improve} \cite{chai2018glaucoma}  \cite{buty2016characterization}\cite{saba2020brain}\end{tabular}}
                                   \\ \cline{3-4}
                                   & & \begin{tabular}[c]{@{}l@{}}Input level fusion\end{tabular} & 
                                   {\begin{tabular}[c]{@{}l@{}}\cite{yang2019dscgans} \cite{xie2019knowledge-based}  \cite{tan2019expert}\cite{liu2019automated}
                                   \cite{feng2020knowledge}\end{tabular}}
                                   \\ \cline{3-4}
                                   & &\begin{tabular}[c]{@{}l@{}}As labels of CNNs\end{tabular} & 
                                   {\begin{tabular}[c]{@{}l@{}}\cite{chen2016automatic}\cite{hussein2017risk} \cite{murthy2017center}\end{tabular}}                          \\ \cline{2-4}
                                 & {\begin{tabular}[c]{@{}l@{}}Other related information
                                 \end{tabular}}& {\begin{tabular}[c]{@{}l@{}}Multi-task learning \\ /network design\end{tabular}}                                                                                                                                                          &\begin{tabular}[c]{@{}l@{}}\cite{liu2018integrate} \cite{wang2018tienet}  \cite{zhang2017tandemnet}\cite{wu2019deep}
                                 \cite{yu2020difficulty}\end{tabular}
                                 \\ \hline
\end{tabular}
\label{tab:diagnosis_knowledge}
\end{table*}


\subsection{Incorporating Knowledge from Natural Datasets or Other Medical Datasets}
\label{sec:diag_natural}


Despite the disparity between natural and medical images, it has been demonstrated that CNNs comprehensively trained on the large-scale well-annotated natural image datasets can still be helpful for disease diagnosis tasks \cite{wimmer2016convolutional}. Intrinsically speaking, this transfer learning process introduces knowledge from natural images into the network for medical image diagnosis.

\begin{figure}[htp!] \begin{centering}
   \includegraphics[width=1\linewidth]{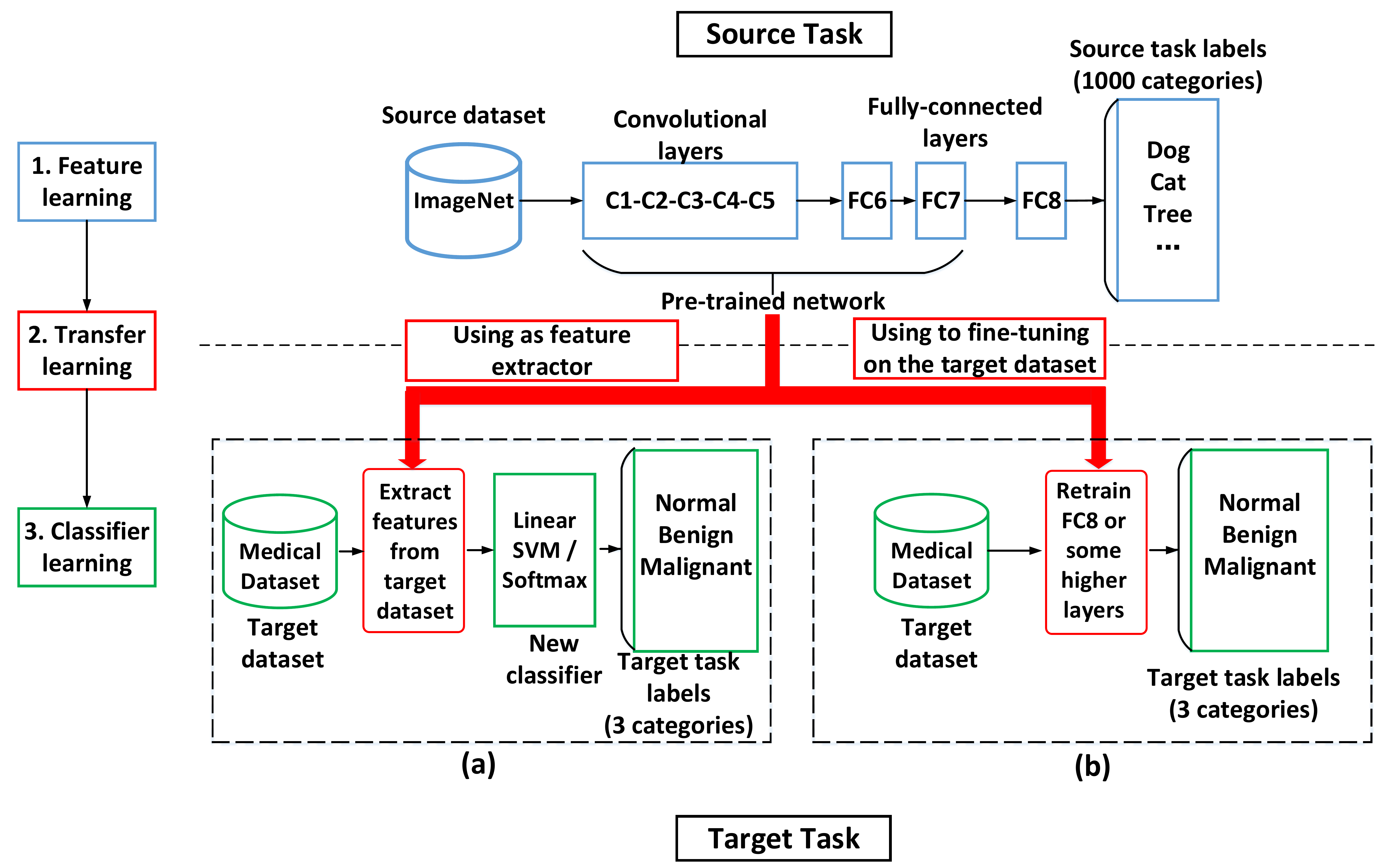}
   \centering
   \caption{Two strategies to utilize the pre-trained network on natural images: (a) as a feature extractor and (b) as an initialization which will be fine-tuned on the target dataset. }
   \label{fig:transfer_learning}
   \end{centering}
 \end{figure}

According to \cite{litjens2017survey}, the networks pre-trained on natural images can be leveraged via two different ways: by utilizing them as fixed feature extractors, and as an initialization which will then be fine-tuned on target medical datasets. These two strategies are illustrated in Fig. \ref{fig:transfer_learning}(a) and Fig. \ref{fig:transfer_learning}(b), respectively.

The first strategy takes a pre-trained network, removes its last fully-connected layer, and treats the rest of the network as a fixed feature extractor. Extracted features are then fed into a linear classifier (e.g., support vector machine (SVM)),  which is trained on the target medical datasets. Applications in this category include mammography mass lesion classification \cite{huynh2016digital} and chest pathology identification \cite{bar2015deep}.

The success of leveraging information from natural images for disease diagnosis can be attributed to the fact that a network pre-trained on natural images, especially in the earlier layers, contain more generic features (e.g., edge detectors and color blob detectors) \cite{yosinski2014transferable}.

In the second strategy, the weights of the pre-trained network are fine-tuned based on the medical datasets. It is possible to fine-tune the weights of all layers in the network, or to keep some of the earlier layers fixed and only fine-tune some higher-level portion of the network. This can be applied to the classification of skin cancer \cite{esteva2017dermatologist}, breast cancer \cite{cao2018improve}, thorax diseases \cite{wang2017chestx}, prostate cancer \cite{li2018path} and interstitial lung diseases \cite{hussein2017risk} .

Besides the information from natural images, using images from other medical datasets is also quite popular.

Medical datasets containing images of the same or similar modality as target images  have similar distribution and therefore can be helpful.
For example, to classify malignant and benign breast masses in digitized screen-film mammograms (SFMs), a multi-task transfer learning DCNN is proposed to incorporate the information from digital mammograms (DMs) \cite{samala2017multi}. It is found to have significantly higher performance compared to the single-task transfer learning DCNN which only utilizes SFMs.




In addition, even medical images with different modalities can provide complementary information. For example, \cite{samala2018breast} uses a model pre-trained on a mammography dataset to show that it could obtain better results than models trained solely on the target dataset comprising digital breast tomosynthesis (DBT) images.
Another example is in prostate cancer classification, where the radiofrequency ultrasound images are first used to train the DCNN, then the model is fine-tuned on B-mode ultrasound images \cite{azizi2017transfer}. Other examples of using the images from different modalities
can be found in \cite{hadad2017classification,li2020digital,han2020deep}.



Furthermore,
as datasets of different classes can help each other in classification tasks \cite{xiao2020pk}, medical datasets featuring images of a variety of diseases can also have similar morphological structures or distribution,
which may be beneficial for other tasks.
For example, a multi-task deep learning (MTDL) method is proposed in \cite{liao2019multi}. 
MTDL can simultaneously utilize multiple cancer datasets so that hidden representations among these datasets can provide more information to small-scale cancer datasets, and enhance the classification performance.
Another example is a cross-disease attention network (CANet) proposed in \cite{li2019canet}. CANet  characterizes and leverages the relationship between diabetic retinopathy (DR) and diabetic macular edema (DME) in fundus images using a special designed disease-dependent attention module. Experimental results on two public datasets show that CANet outperforms other methods on diagnosing both of the two diseases.

\subsection{Incorporating Knowledge from Medical Doctors}
\label{sec:diag_rediologists}

Experienced medical doctors can give fairly accurate conclusion on the given medical images, mainly thanks to the training they have received and the expertise they have accumulated over many years. In general, they often follow some certain patterns or take some procedures when reading medical images. Incorporating these types of knowledge can improve the diagnostic performance of deep learning models.

The types of medical domain knowledge utilized in deep learning models for disease diagnosis can be summarized into the following five categories:
\begin{enumerate}
\item the training pattern,
\item the general diagnostic patterns they view images,
\item the areas on which they usually focus,
\item the features (e.g., characteristics, structures, shapes) they give special attention to, and
\item other related information for diagnosis.
\end{enumerate}
The research works for each category will be described in the following sections.


\subsubsection{Training Pattern of Medical Doctors}
\label{sec:diag_trainingprocess}


The training process of medical students has a character: they are trained by tasks with increasing difficulty.
For example, students begin with some easier tasks, such as deciding whether an image contains lesions, later are required to accomplish more challenging tasks, such as determining whether the lesions are benign or malignant. Over time, they will advance to more difficult tasks, such as determining the subtypes of lesions in images.

This training pattern can be introduced in the training process of deep neural networks via curriculum learning \cite{bengio2009curriculum}. Curriculum determines a sequence of training samples ranked in ascending order of learning difficulty. Curriculum learning has been an active research topic in computer vision and has been recently utilized for medical image diagnosis. 

For example, a teacher-student curriculum learning strategy is proposed for breast screening classification from DCE-MRI \cite{maicas2018training}. The deep learning model is trained on simpler tasks before introducing the hard problem of malignancy detection. This strategy shows the better performance when compared with the other methods.

Similarly, \cite{tang2018attention} presents a CNN based attention-guided curriculum learning framework by leveraging the severity-level attributes mined from radiology reports. Images in order of difficulty (grouped by different severity-levels) are fed into the CNN to boost the learning process gradually.

In \cite{jimenez2019medical}, the curriculum learning is adopted to support the classification of proximal femur fracture from X-ray images.  The approach assigns a degree of difficulty to each training sample. By first learning `easy' examples and then `hard' ones, the model can reach a better performance even with fewer data. Other examples of using curriculum learning for disease diagnosis can be found in \cite{haarburger2019multi,zhao2020egdcl,jimenez2020curriculum,wei2020learn,qi2020curriculum}.





\subsubsection{General Diagnostic Patterns of Medical Doctors}
\label{sec:diag_diagnosticpattern}

Experienced medical doctors generally follow some patterns when they read medical images. These patterns can be integrated into deep learning models with appropriately modified architectures.

\begin{figure}[htp!] \begin{centering}
   \includegraphics[width=1.0\linewidth]{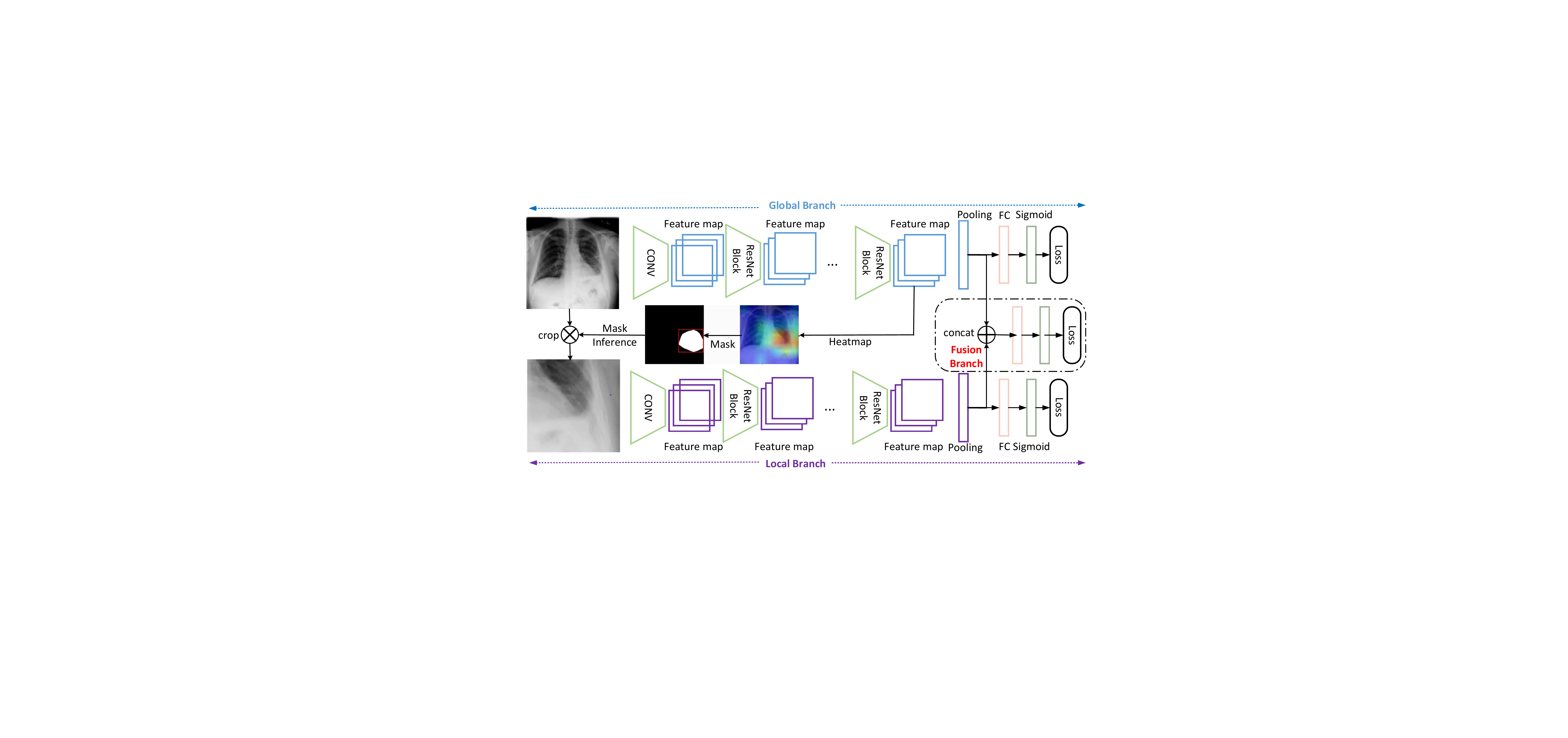}
   \centering
   \caption{The example of leveraging the diagnostic pattern of radiologists for thorax disease diagnosis \cite{Diagnose}, where three branches are used to extract and combine the global and local features.}
    \label{fig:diagnosepattern}
   \end{centering}

 \end{figure}

For example, radiologists generally follow a three-staged approach when they read chest X-ray images: first browsing the whole image, then concentrating on the local lesion areas,  and finally combining the global and local information to make decisions. This pattern is incorporated  in the architecture design of the network for thorax disease classification \cite{Diagnose} (see Fig. \ref{fig:diagnosepattern}). The proposed network has three  branches, one is used to view the whole image, the second for viewing the local areas, and the third one for combining the global and local information together. The network yields state-of-the-art accuracy on the ChestX-ray14 dataset. In addition, besides the information from the whole image and local lesion area, the information from lung area is also leveraged in \cite{wang2020learning}. In particular,  a segmentation subnetwork is first used to locate the lung area from the whole image, and then lesion areas are generated by using an attention heatmap. Finally, the most discriminative features are fused for final disease prediction. Another example is a Dual-Ray Net proposed to deal with the front and lateral chest radiography simultaneously \cite{huang2020dual}, which also mimics the reading pattern of radiologists.

In the diagnosis of skin lesions, experienced dermatologists generally first locate lesions, then identify dermoscopic features from the lesion areas, and finally make diagnosis based on the features.
This pattern is mimicked in the design of the network for the diagnosis of skin lesions \cite{gonzalez2018dermaknet}. The proposed network, DermaKNet, comprised  several subnetworks with dedicated tasks: lesion-skin segmentation, detection of dermoscopic features, and global lesion diagnosis. The DermaKNet achieves higher performance compared to the traditional CNN models.

In addition, in mass identification in mammogram, radiologists generally analyze the bilateral and ipsilateral views simultaneously. To emulate this reading practice, \cite{yang2020momminet} proposes MommiNet to perform end-to-end bilateral and ipsilateral analysis of mammogram images. In addition, symmetry and geometry constraints are also aggregated from these views. Experiments show the state-of-the-art mass identification  accuracy on DDSM. Another example of leveraging this diagnostic pattern of medical doctors can be found in skin lesion diagnosis and thorax disease classification \cite{liu2020semi}.

\subsubsection{The Areas Medical Doctors Usually Focus on}
\label{sec:diag_attentionmap}
When experienced medical doctors read images, they generally focus on a few specific areas, as these areas are more informative than other places for the purpose of disease diagnosis. Therefore, the information about where medical doctors focus may help deep learning models yield better results.



\begin{figure}[!htp] \begin{centering}
   \includegraphics[width=0.8\linewidth]{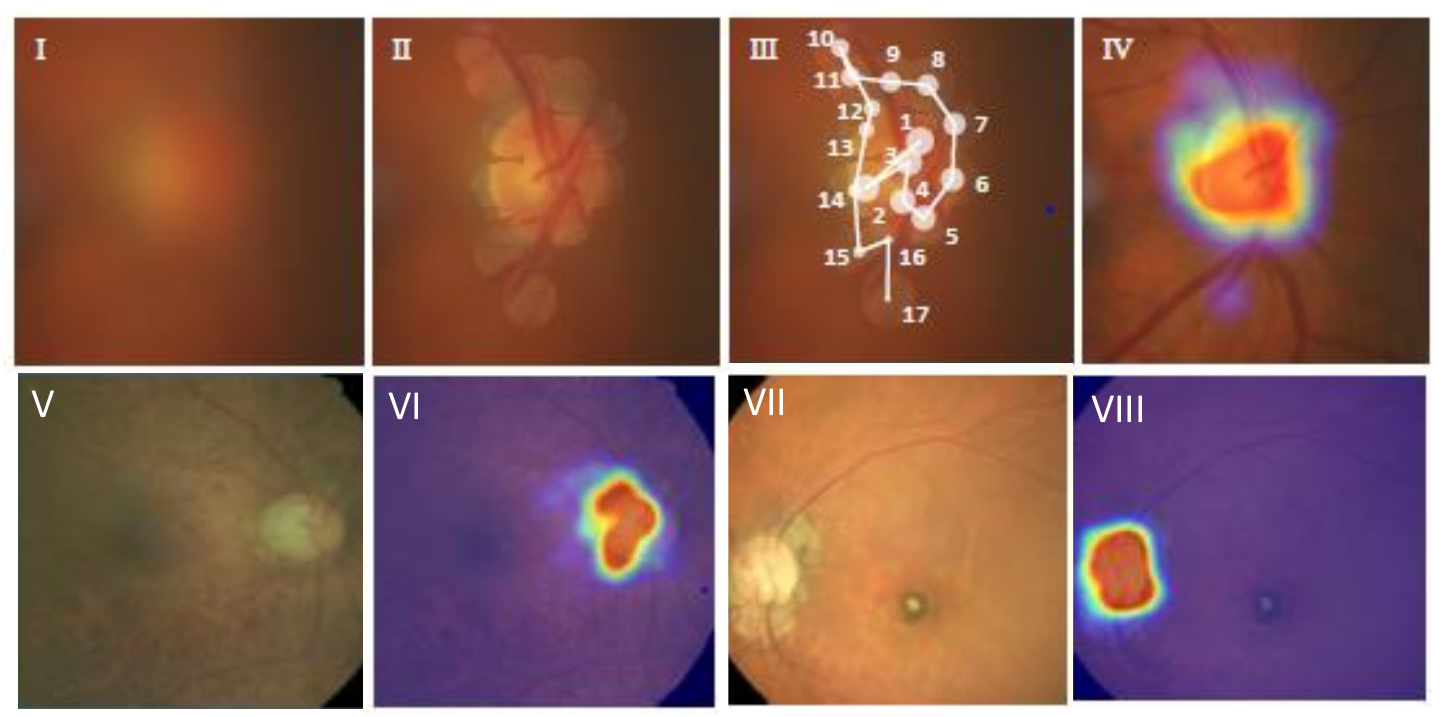}
   \centering
   \caption{Example of capturing the attention maps of an ophthalmologist in glaucoma diagnosis \cite{li2019attention}. I, II, III and IV are the original blurred fundus image, the fixations of ophthalmologists with cleared regions, the  order of clearing the blurred regions, and the generated attention map based on the captured fixations, respectively. V and VII represent the original fundus images. VI and VIII are the corresponding attention maps of V and VII generated by using the method in I-IV.}
   \label{fig:attentionmap}
   \end{centering}
 \end{figure}

The knowledge above is generally represented as `attention maps', which are annotations given by medical doctors indicating the areas they focus on when reading images. For example, a CNN named AG-CNN explicitly incorporates the `attention maps' for glaucoma diagnosis \cite{li2019attention}. The attention maps of ophthalmologists are collected through a simulated eye-tracking experiment, which are used to indicate where they focus when reading images. An example of capturing the attention maps of an ophthalmologist in glaucoma diagnosis is shown in Fig. \ref{fig:attentionmap}. To incorporate the attention maps,  an attention prediction subnet in AG-CNN is designed, and the attention prediction loss measuring the difference between the generated and ground truth attention maps (provided by ophthalmologists) is utilized to supervise the training process. Experimental results show that AG-CNN significantly outperforms
the state-of-the-art glaucoma detection methods.

Another example in this category is the lesion-aware CNN (LACNN) for the classification of retinal optical coherence tomography (OCT) images \cite{8637959}. The LACNN simulates the pattern of ophthalmologists' diagnosis by focusing on local lesion-related regions. Concretely, the `attention maps' are firstly represented as the annotated OCT images delineating the lesion regions using bounding polygons. To incorporate the information, the LACNN proposes a lesion-attention module to enhance the features from local lesion-related regions while still preserving the meaningful structures in global OCT images. Experimental results on two clinically acquired OCT datasets demonstrate the effectiveness of introducing attention maps for retinal OCT image classification, with 8.3\% performance gain when compared with the baseline method.


Furthermore, \cite{mitsuhara2019embedding} proposes an Attention
Branch Network (ABN) to incorporate the knowledge given by the radiologists in diabetic retinopathy. ABN introduces a branch structure which generates attention maps that highlight the attention regions of the network. During the training process, ABN allows the attention maps to be modified with semantic segmentation labels of disease regions. The semantic labels are also annotated by radiologists as the ground truth attention maps. Experimental results on the disease grade recognition of retina images show that ABN achieves 93.73\%  classification accuracy and its interpretability is clearer than conventional approaches.

Other examples of incorporating attention maps of medical doctors can be found in the diagnosis of radiotherapy-related esophageal fistula \cite{cui2020collaborative}, breast cancer diagnosis \cite{xiedg}, and short-term lesion change detection in melanoma screening \cite{zhang2020short}.

\subsubsection{Features That Medical Doctors Give Special Attention to}
\label{sec:diag_diseases}

In the last decades, many guidelines and rules have gradually developed in various medical fields to point out some important features for diagnosis. These features are called \emph{`hand-crafted features'} as they are designated by medical doctors.  For example, the popular ABCD rule \cite{nachbar1994abcd} is widely used by dermatologists to classify melanocytic tumors. The ABCD rule points out four distinguishing features, namely asymmetry, border, color and differential structures, to determine whether a melanocytic skin lesion under the investigation is benign or malignant.  

\begin{table}[htp!]
\footnotesize
\center
\caption{Features in the BI-RADS guideline to classify benign and malignant breast tumors in ultrasound images \cite{birads}}
\begin{tabular}{|c|c|c|}
\hline
& Benign  & Malignant\\
\hline
Margin & smooth, thin, regular  & irregular, thick \\
\hline
Shape & round or oval &irregular                         \\
\hline
Microcalcification                   & no                        & yes                               \\
\hline
Echo Pattern                         & clear & unclear \\
\hline
Acoustic Attenuation                 & not obvious               & obvious                           \\
\hline
Side Acoustic Shadow                 & obvious                   & not obvious                       \\ 
\hline
\end{tabular}
\label{tab:1}
\end{table}

\begin{figure}[htp!]
\begin{centering}
   \includegraphics[width=0.9\linewidth]{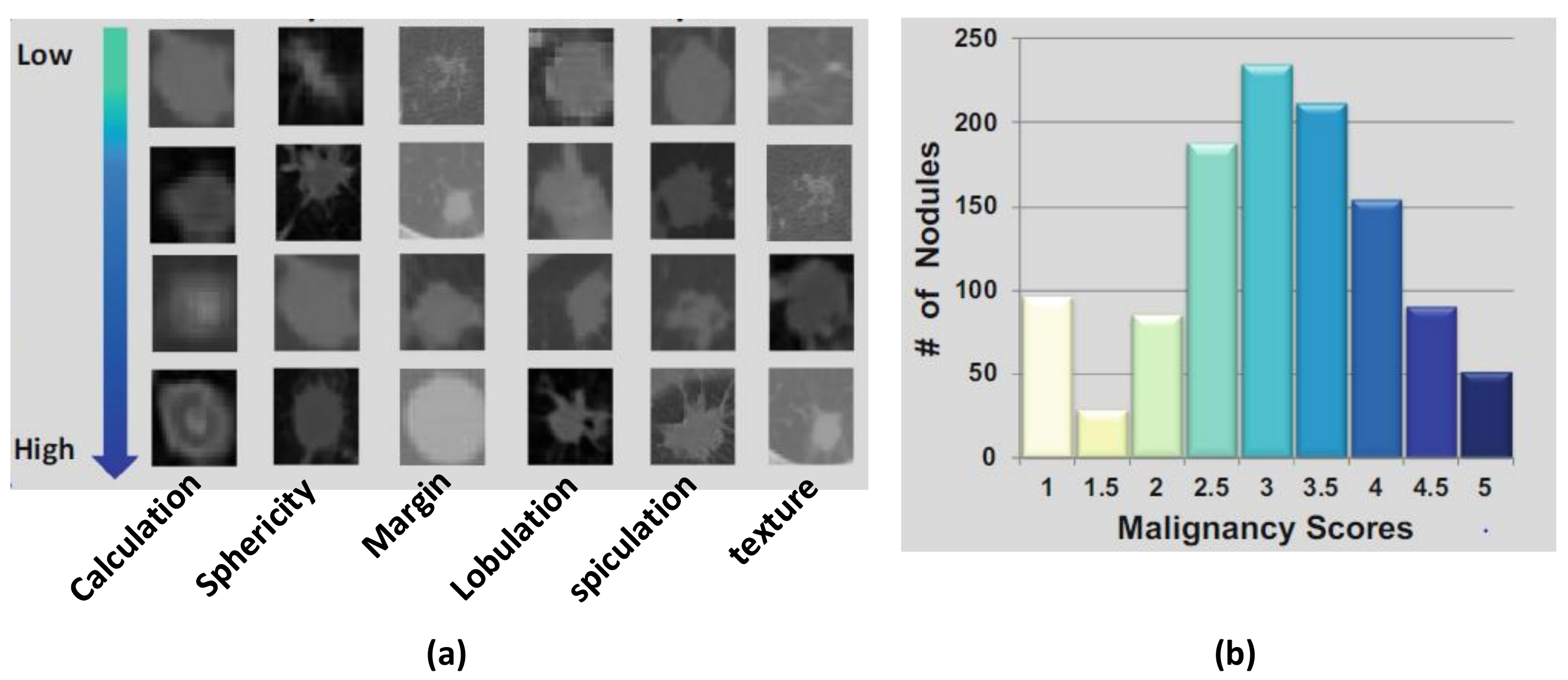}
   \centering
   \caption{Lung nodule attributes with different malignancy scores \cite{hussein2017risk}.(a) From top to the bottom, six different nodule features attribute  from missing to highest prominence. (b) The number of nodules with different malignancy scores.}
   \label{fig:lung_nodule}
   \end{centering}
 \end{figure}

Another example is in the field of breast cancer diagnosis. Radiologists use the BI-RADS  (Breast Imaging Reporting and Data System) score \cite{birads} to place abnormal findings into different categories, with a score of 1 indicating negative findings and a score of 6 indicating breast cancer. More importantly, for each imaging modality, BI-RADS indicates some features closely related to its scores, including margin, shape, micro-calcification, and echo pattern. For example, for breast ultrasound images, tumors  with smooth, thin and regular margins are more likely to be benign, while tumors with irregular and thick margins are highly suspected to be malignant. Other features that can help to classify benign and malignant breast tumors are shown in Table \ref{tab:1}.
Similarly, for the benign-malignant risk assessment of lung nodules in  \cite{hussein2017risk}, six high-level nodule features, including calcification, sphericity, margin, lobulation, spiculation and texture, have shown a tightly connection with malignancy scores (see Fig. \ref{fig:lung_nodule}).

These different kinds of hand-crafted features have been widely used in many traditional CAD systems. These systems generally first extract these features from medical images, and then feed them into some classifiers like SVM or Random Forest \cite{breiman2001random,cortes1995support}. For example, for the lung nodule classification on CT images, many CAD systems utilize features including the size, shape, morphology, and texture from the suspected lesion areas \cite{alilou2017intra,xie2019knowledge-based}. Similarly, in the CAD systems for the diagnosis of breast ultrasound images, features such as intensity, texture and shape are used \cite{hsu2019breast}.

When using deep learning models like CNNs, which have the ability to automatically extract representative features, there are four approaches to combining  `hand-crafted features' with features extracted from CNNs.
\begin{itemize}
\item Decision-level fusion: the two types of features are fed into two classifiers separately, then the decisions from the two classifiers are combined.
\item Feature-level fusion: the two types of features are directly combined via techniques such as concatenation.
\item Input-level fusion: the hand-crafted features are represented as image patches, which are then taken as inputs to the CNNs.
\item 
Usage of features as labels of CNNs: the hand-crafted features are first annotated and then utilized as labels for CNNs during the training process.
\end{itemize}

\begin{figure}[htp!]
\begin{centering}
   \includegraphics[width=1\linewidth]{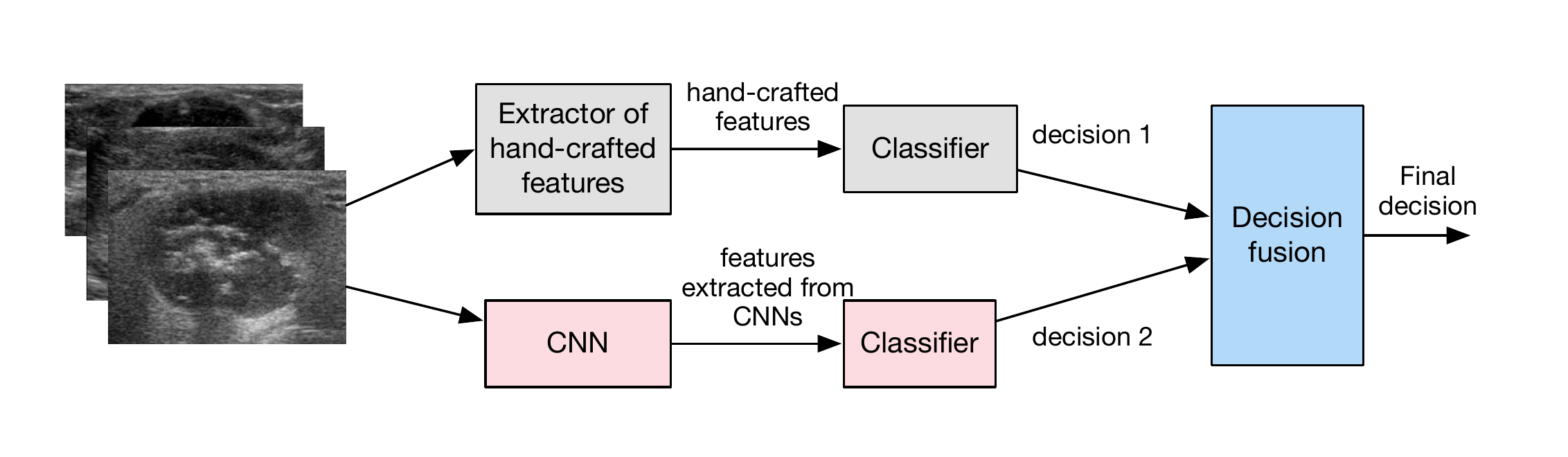}
   \centering
   \caption{Decision-level fusion: the decisions from two classifiers, one based on hand-crafted features, and the other on the CNNs, are combined.}
   \label{fig:decision_level}
   \end{centering}
 \end{figure}

\textbf{Decision-level fusion}:  The structure of this approach is illustrated in Fig. \ref{fig:decision_level}. In this approach, the hand-crafted features and the features extracted from CNNs are separately fed into two classifiers. Then, the classification results from both classifiers are combined using some decision fusion techniques to produce final classification results.

For example, a skin lesion classification model proposed in \cite{majtner2016combining} combines the results from two SVM classifiers. The first one uses hand-crafted features (i.e., RSurf features and local binary patterns (LBP)) as input and the second employs features derived from a CNN. Both of the classifiers predict the category for each tested image with a classification score. These scores are subsequently used to determine the final classification result.

Similarly, a mammographic tumor classification method also combines features in decision-level fusion \cite{huynh2016digital}. After individually performing classification with CNN features and analytically extracted features (e.g., contrast, texture, and margin spiculation), the method adopts the soft voting to combine the outputs from both individual classifiers. Experimental results show that the performance of the ensemble classifier was significantly better than the individual ones. Other examples that utilize this approach include lung nodule diagnosis \cite{xie2018fusing}, breast cancer diagnosis \cite{antropova2017deep}, the classification of cardiac slices \cite{moradi2016hybrid} and the prediction of the invasiveness risk of stage-I lung adenocarcinomas \cite{xia2020comparison}.

\begin{figure}[htp!]
\begin{centering}
\vspace{-0.2cm}
   \includegraphics[width=0.9\linewidth]{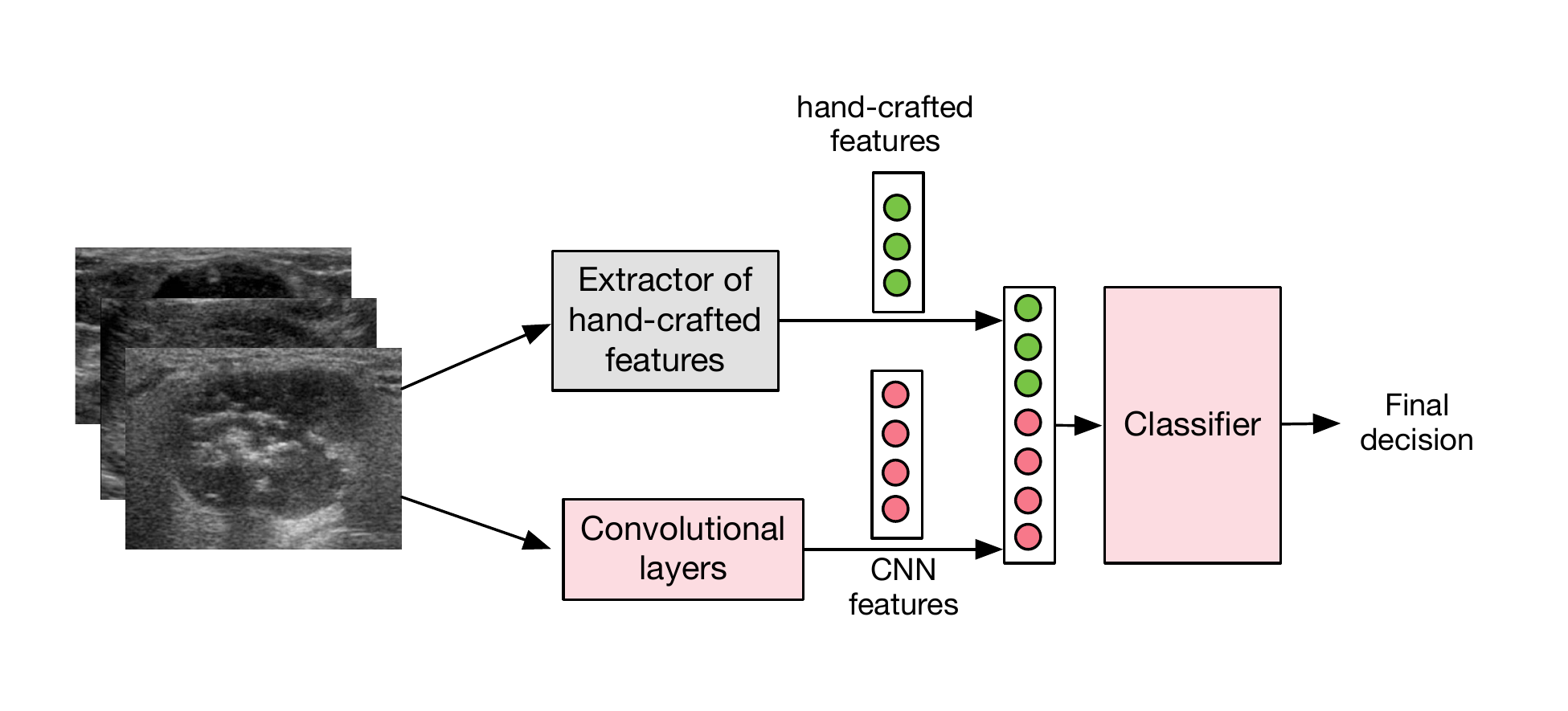}
   \centering
   \caption{Feature-level fusion: the hand-crafted features are combined with the features extracted from CNNs as the new feature vectors.}
   \label{fig:feature_level_fusion}
   \end{centering}
 \end{figure}

\textbf{Feature-level fusion}:
In this approach, hand-crafted features and features extracted from CNNs are concatenated, and the combined features are fed into a classifier for diagnosis.  The structure of this approach is illustrated in Fig. \ref{fig:feature_level_fusion}.

For example, a combined-feature based classification approach called CFBC is proposed for lung nodule classification by \cite{xie2016lung}. In CFBC, the hand-crafted features (including texture and shape descriptors) and the features learned by a nine-layer CNN are combined and fed into a back-propagation neural network. Experimental results derived from a publicly available dataset show that compared with a purely CNN model, incorporating hand-crafted features improves the accuracy, sensitivity, and specificity by 3.87\%, 6.41\%, and 3.21\%, respectively.

Another example in this category is the breast cancer classification in histology images \cite{cao2018improve}. More specifically, two hand-crafted features, namely the parameter-free threshold adjacency statistics and the gray-level co-occurrence matrix, are fused with the five groups of deep learning features extracted from five different deep models. The results show that after incorporating hand-crafted features, the accuracy of the deep learning model can be significantly improved.


Other examples of employing feature-level fusion can be found in glaucoma diagnosis \cite{chai2018glaucoma}, skin lesion classification \cite{hagerty2019deep}, lung nodule classification \cite{buty2016characterization} and brain tumor diagnosis \cite{saba2020brain}. 

\begin{figure}[htp!]
\begin{centering}
   \includegraphics[width=0.9\linewidth]{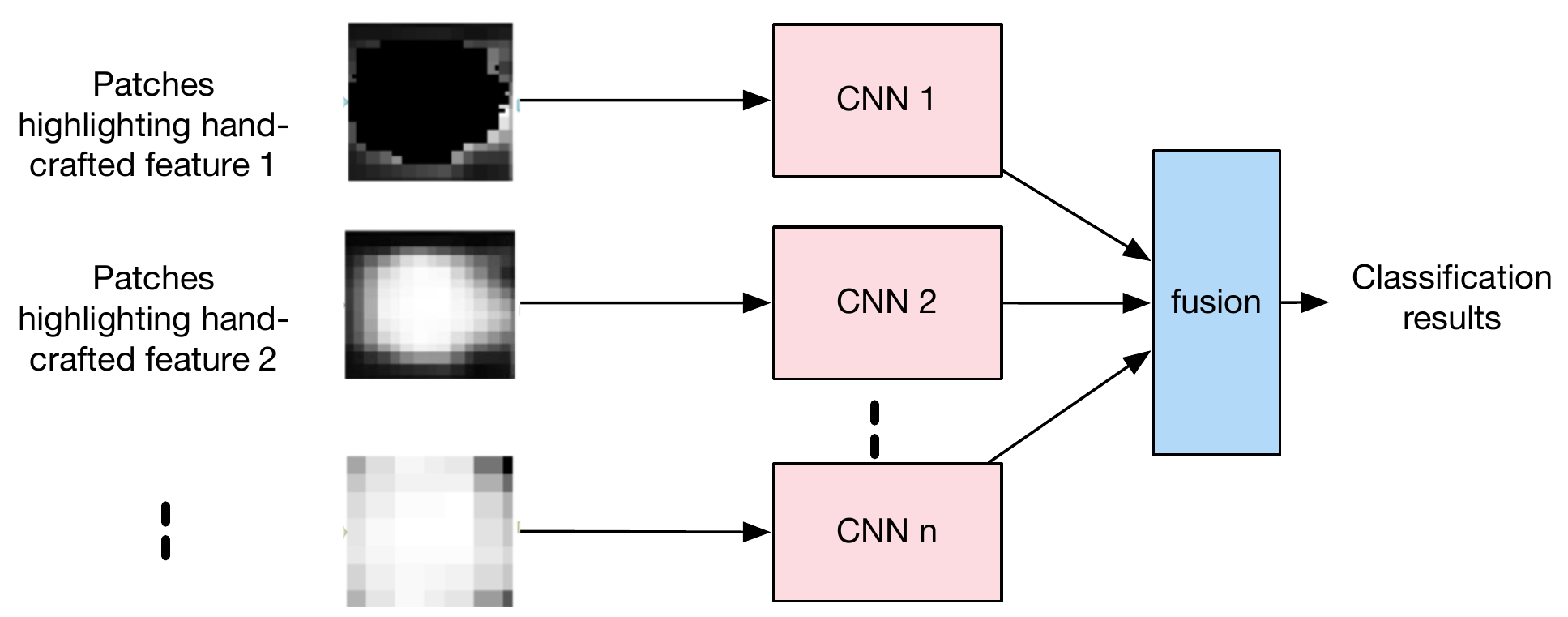}
   \centering
   \caption{Input-level fusion: the hand-crafted features are represented as image patches that are taken as inputs to the CNNs.}
   \label{fig:input_level_fusion}
   \end{centering}
 \end{figure}

\textbf{Input-level fusion}:
In this approach, hand-crafted features are first represented as patches highlighting the corresponding features. Then, these patches are taken as input to CNNs to make the final conclusion. Using this approach, the CNNs are expected to rely more on the hand-crafted features. It should be noted that generally speaking, one CNN is required for each type of hand-crafted feature, and information obtained from these CNNs need to be combined in some manner to make a final decision. The structure of this approach is illustrated in Fig. \ref{fig:input_level_fusion}. 

For example, in \cite{tan2019expert}, three types of hand-crafted features, namely the contrast information of the initial nodule candidates (INCs) and the outer environments, the histogram of oriented gradients (HOG) feature, and the LBP feature, are transformed into the corresponding patches and are taken as inputs of multiple CNNs. The results show that this approach outperforms both conventional CNN-based approaches and traditional
machine-learning approaches based on hand-crafted features.

Another example using input-level fusion approach is the MV-KBC (multi-view knowledge-based collaborative) algorithm proposed for lung nodule classification  \cite{xie2019knowledge-based}. The MV-KBC first extracts patches corresponding to three features: the overall appearance (OA), nodule's heterogeneity in voxel values (HVV) and heterogeneity in shapes (HS). Each type of patch is fed into a ResNet. Then, the outputs of these ResNets are combined to generate the final classification results.

Moreover, \cite{yang2019dscgans} proposes the dual-path semi-supervised conditional generative adversarial networks (DScGAN) for thyroid classification. More specifically, the features from the ultrasound B-mode images and the ultrasound elastography images are first extracted as the OB patches (indicating the region of interest (ROI) in B-mode images), OS patches (characterizing the shape of the nodules), and OE patches (indicating the elasticity ROI according to the B-mode ROI position). Then, these patches are used as the input of the DScGAN. Using the information from these patches is demonstrated to be effective to improve the classification performance. Other examples employ input-level fusion can be found in thyroid nodule diagnosis \cite{liu2019automated} and breast cancer diagnosis on multi-sequence MRI \cite{feng2020knowledge}.

\textbf{Usage of features as labels of CNNs:} In this approach, besides the original classification labels of images, medical doctors also provide labels for some hand-crafted features. This extra information is generally incorporated into deep learning models via the multi-task learning structure.

For example, in \cite{chen2016automatic}, the nodules in lung images were quantitatively rated by radiologists with regard to nine hand-crafted features (e.g., spiculation, texture, and margin). The multi-task learning framework is used to incorporate the above information into the main task of lung nodule classification.

In addition, for the benign-malignant risk assessment of lung nodules in low-dose CT scans \cite{hussein2017risk}, the binary labels about the presence of six high-level nodule attributes, namely the calcification, sphericity, margin, lobulation, spiculation and texture, are utilized. Six CNNs are designed and each one aims at learning one attribute.  The discriminative features automatically learned by CNNs for these attributes are fused in a multi-task learning framework to obtain the final risk assessment scores.

Similarly, in \cite{murthy2017center}, each glioma nuclear image is exclusively labeled in terms of the shapes and attributes for the centermost nuclei of the image. These features are then learned by a multi-task CNN. Experiments demonstrate that the proposed method outperforms the baseline CNN.

\subsubsection{Other Types of Information Related to Diagnosis}
\label{sec:diag_extraannotations}

In this section, we summarize other types of information that can help deep learning models to improve their diagnostic performance. These types of information include extra category labels and clinical diagnostic reports.


\textbf{Extra category labels}

For medical images, besides a classification label (i.e., normal, malignant or benign), radiologists may give some extra category labels. For example, in the ultrasonic diagnosis of breast cancer, an image usually has a BI-RADS label that  classifies the image into 0$\sim$6 \cite{birads}---category 0 suggests re-examination, categories 1 and 2 indicate that it is prone to be a benign lesion, category 3 suggests probably benign findings, categories 4 and 5 are suspected to be malignant, category 6 definitely suggests malignant).
These labels also contain information about the condition of lesions. In \cite{liu2018integrate}, the BI-RADS label for each medical image is incorporated in a multi-task learning structure as the label of an auxiliary task. Experimental results show that incorporating these BI-RADS labels can improve the diagnostic performance of major classification task. Another example of using the information from BI-RADS labels can be found in \cite{wu2019deep}.

In addition, during the process of image annotation, consensus or disagreement among experts regarding images reflects the gradeability and difficulty levels of the image, which is also a  representation of medical domain knowledge. To incorporate this information,\cite{yu2020difficulty} proposes a multi-branch model structure to predict the most sensitive, most specifical and a balanced fused result for glaucoma images. Meanwhile, the  consensus loss is also used to encourage the sensitivity and specificity branch to generate consistent and contradictory predictions for images with consensus and disagreement labels, respectively. 


\textbf{Extra clinical diagnostic reports}

A clinical report is a summary of all the clinical findings and impressions determined during examination of a radiography study.
It usually contains rich information and reflects the findings and descriptions of medical doctors. Incorporating clinical reports into CNNs designed for disease diagnosis is typically beneficial. As medical reports are generally handled by recurrent neural networks (RNNs), to incorporate information from medical reports, generally hybrid networks containing both CNNs and RNNs are needed.

For example, the Text-Image embedding network (TieNet) is designed to classify the common thorax disease in chest X-rays \cite{wang2018tienet}. TieNet has an end-to-end CNN-RNN architecture enabling it to integrate information of radiological reports. The classification results are significantly improved (with about a 6\% increase on average in AUCs) compared with the baseline CNN purely based on medical images.

In addition, using semantic information from diagnostic reports is explored in \cite{zhang2017tandemnet} for understanding pathological bladder cancer images. A dual-attention model is designed to facilitate the high-level interaction of semantic information and visual information. Experiments demonstrate that incorporating information from diagnostic reports significantly improves the cancer diagnostic performance over the baseline method. 
\subsection{Summary}
\label{sec:diag_overview}
In the previous sections, we introduced different kinds of domain knowledge and the corresponding integrating methods into the deep learning models for disease diagnosis. Generally, almost all types of domain knowledge are proven to be effective in boosting the diagnostic performance, especially using the metrics of accuracy and AUC, some examples and their quantitative improvements are listed in Table \ref{tab:diag_quan}.

\begin{table*}[]
\small
\caption{The comparison of the quantitative metrics for some disease diagnosis methods after integrating domain knowledge}
\center
\begin{tabular}{|l|l|l|l|}
\cline{1-4}
\multicolumn{1}{|c|}{Reference}                              & \multicolumn{1}{c|}{\begin{tabular}[c]{@{}l@{}}Baseline Model/With Domain Knowledge\end{tabular}}                              & \multicolumn{1}{c|}{Accuracy}   & \multicolumn{1}{c|}{AUC}           \\ \hline
\multirow{1}{*}{\cite{Diagnose}} & \multicolumn{1}{c|}{\begin{tabular}[c]{@{}l@{}}AG-CNN only with\\ global branch/AG-CNN\end{tabular}} & \multicolumn{1}{c|}{--/--}                 & \multicolumn{1}{c|}{0.840/0.871}     \\ \hline
\multirow{1}{*}{\cite{gonzalez2018dermaknet}} & \multicolumn{1}{c|}{\begin{tabular}[c]{@{}l@{}}ResNet-50/DermaKNet\end{tabular}} & \multicolumn{1}{c|}{--/--}                  & \multicolumn{1}{c|}{0.874/0.908}    \\ \hline 
\multirow{1}{*}{\cite{hadad2017classification}} & \multicolumn{1}{c|}{\begin{tabular}[c]{@{}l@{}}Fine-tuned VGG-Net/\\Fine-tuned MG-Net\end{tabular}} & \multicolumn{1}{c|}{0.900/0.930}             & \multicolumn{1}{c|}{0.950/0.970}  \\ \hline
\multirow{1}{*}{\cite{hagerty2019deep}} & \multicolumn{1}{c|}{\begin{tabular}[c]{@{}l@{}}ResNet-50/ResNet-50 \\with handcrafted features\end{tabular}}  &\multicolumn{1}{c|}{--/--}               & \multicolumn{1}{c|}{0.830/0.940}  \\ \hline
\multirow{1}{*}{\cite{li2019attention}} & \multicolumn{1}{c|}{\begin{tabular}[c]{@{}l@{}}CNN without using \\attention map/AG-CNN\end{tabular}}    & \multicolumn{1}{c|}{0.908/0.953}            & \multicolumn{1}{c|}{0.966/0.975} \\ \hline
\multirow{1}{*}{\cite{li2019canet}} &  \multicolumn{1}{c|}{\begin{tabular}[c]{@{}l@{}}ResNet50/CANet\end{tabular}}      & \multicolumn{1}{c|}{0.828/0.851}            & \multicolumn{1}{c|}{--/--}  \\ \hline
\multirow{1}{*}{\cite{liu2018integrate}} &  \multicolumn{1}{c|}{\begin{tabular}[c]{@{}l@{}}VGG16/Multi-task VGG16\end{tabular}}     & 0.829 /0.833           & \multicolumn{1}{c|}{--/--}   \\ \hline
\multirow{1}{*}{\cite{maicas2018training}} &  \multicolumn{1}{c|}{DenseNet/BMSL}  & \multicolumn{1}{c|}{--/--}                 & \multicolumn{1}{c|}{0.850/0.900}   \\ \hline
\multirow{1}{*}{\cite{samala2018breast}} &  \multicolumn{1}{c|}{\begin{tabular}[c]{@{}l@{}}CNN with single-stage/\\multi-stage transfer learning\end{tabular}}  &  \multicolumn{1}{c|}{--/--}                & \multicolumn{1}{c|}{0.850/0.910}  \\ \hline
\multirow{1}{*}{\cite{tang2018attention}} &  \multicolumn{1}{c|}{\begin{tabular}[c]{@{}l@{}}CNN/AGCL\end{tabular}}      & \multicolumn{1}{c|}{--/--}                 & \multicolumn{1}{c|}{0.771/0.803}   \\ \hline
\multirow{1}{*}{\cite{yang2019dscgans}} &  \multicolumn{1}{c|}{\begin{tabular}[c]{@{}l@{}}DScGAN without/with \\ domain knowledge\end{tabular}}  & \multicolumn{1}{c|}{0.816/0.905}        & 0.812/0.914   \\ \hline
\end{tabular}
\label{tab:diag_quan}
\end{table*}

With respect to type of domain knowledge for disease diagnosis, the knowledge from natural images is widely incorporated in deep learning models.  However, considering the gap between natural images and medical ones, using information from external medical datasets of the same diseases with similar modalities (e.g., SFM and DM)  \cite{samala2017multi}, with different modalities (DBT and MM, Ultrasound) \cite{samala2018breast}, and even with different diseases  \cite{liao2019multi} may be more effective. To incorporate the above information is relatively easy, and both transfer learning and multi-task learning have been widely adopted.  However, there are still no comparative studies on how different extra datasets can improve the performance of deep learning models.

For the domain knowledge of medical doctors, the high-level domain knowledge (e.g., the training pattern and the diagnostic pattern) is generally common for different diseases. On the other hand, the low-level domain knowledge, such as the areas in images and features of diseases that medical doctors usually pay attention to, is generally different for different diseases. There is generally a trade-off between the versatility and the utility of domain knowledge. To diagnose a certain disease, the benefit of incorporating a versatile type of domain knowledge suitable for many diseases may not be as significant as using the one that is specific for the disease. However, identifying such specific domain knowledge may not be easy, and generally requires more efforts from medical doctors (e.g., to identify hand-crafted features or attention maps).

We believe that more kinds of medical domain knowledge can be explored and utilized for disease diagnosis. In addition, comparative studies on some benchmark datasets should be carried out with respect to different types of domain knowledge and different incorporating methods for disease diagnosis. The results can provide further insights about the utility of medical domain knowledge for deep learning models.

\section{Lesion, Organ, and Abnormality Detection}
\label{sec:detection}
Detecting lesions in medical images is an important task and a key part for disease diagnosis in many conditions. Similarly, organ detection is an essential preprocessing step for image registration, organ segmentation, and lesion detection. Detecting abnormalities in medical images, such as cerebral microbleeds in brain MRI images and hard exudates in retinal images, is also required in many applications.

In this section, the deep learning models widely used for object detection in medical images are first described in Subsection \ref{sec:detmodel}. Then, the existing works on utilizing domain knowledge from natural and medical datasets, and from medical doctors are introduced in detail in Subsection \ref{sec:det_natural} and Subsection \ref{sec:det_radiologists}, respectively. Lastly, we summarize and discuss these different types of domain knowledge and the associated incorporating methods in Subsection \ref{sec:det_overview}.

\subsection{General Structures of Deep Learning Models for Object Detection in Medical Images}
\label{sec:detmodel}

Deep learning models designed for object detection in natural images have been directly applied to detect objects in medical images. These applications include  pulmonary nodule detection in CT images \cite{setio2016pulmonary}, retinal diseases detection in retinal fundus \cite{gulshan2016development} and so on.

 \begin{figure}[!htp] \begin{centering}
    \includegraphics[width=0.9\linewidth]{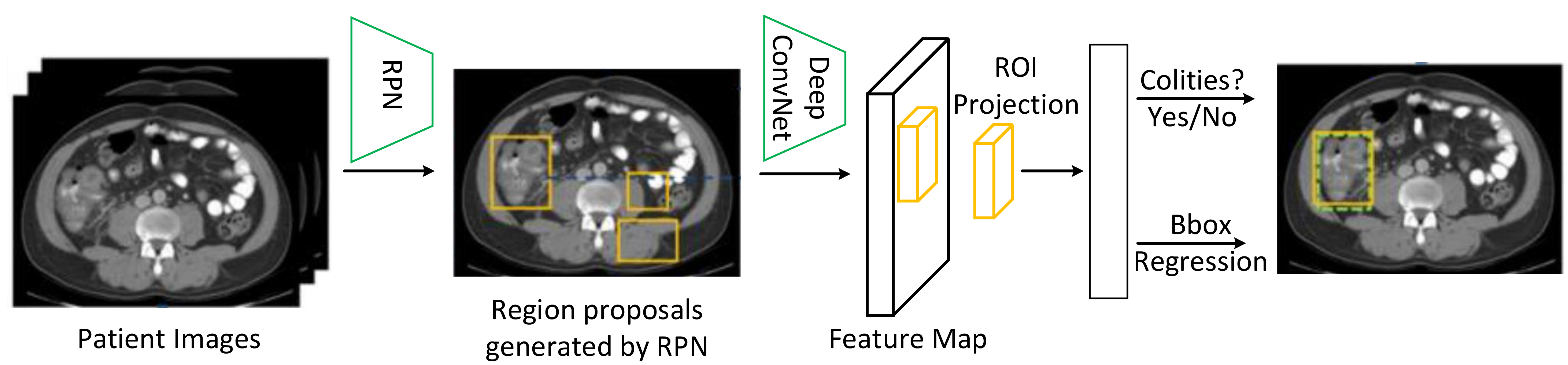}
    \centering
    \caption{
    The workflow of colitis detection method by using the Faster R-CNN structure  \cite{liu2017detection}.
    }
    \label{fig:det_structure2}
    \end{centering}
  \end{figure}

One popular type of model is the two-stage detectors like the Faster R-CNN \cite{ren2015faster} and the Mask R-CNN \cite{he2017mask}. They generally consist of a region proposal network (RPN) that hypothesizes candidate object locations and a detection network that refines region proposals. Applications in this category include colitis detection in CT images \cite{liu2017detection}, intervertebral disc detection in X-ray images \cite{sa2017intervertebral} and the detection of architectural distortions in mammograms \cite{ben2017domain}. Fig. \ref{fig:det_structure2} shows an example of using Faster R-CNN structure for colitis detection \cite{liu2017detection}.

\begin{figure}[!htp] \begin{centering}
    \includegraphics[width=1.0\linewidth]{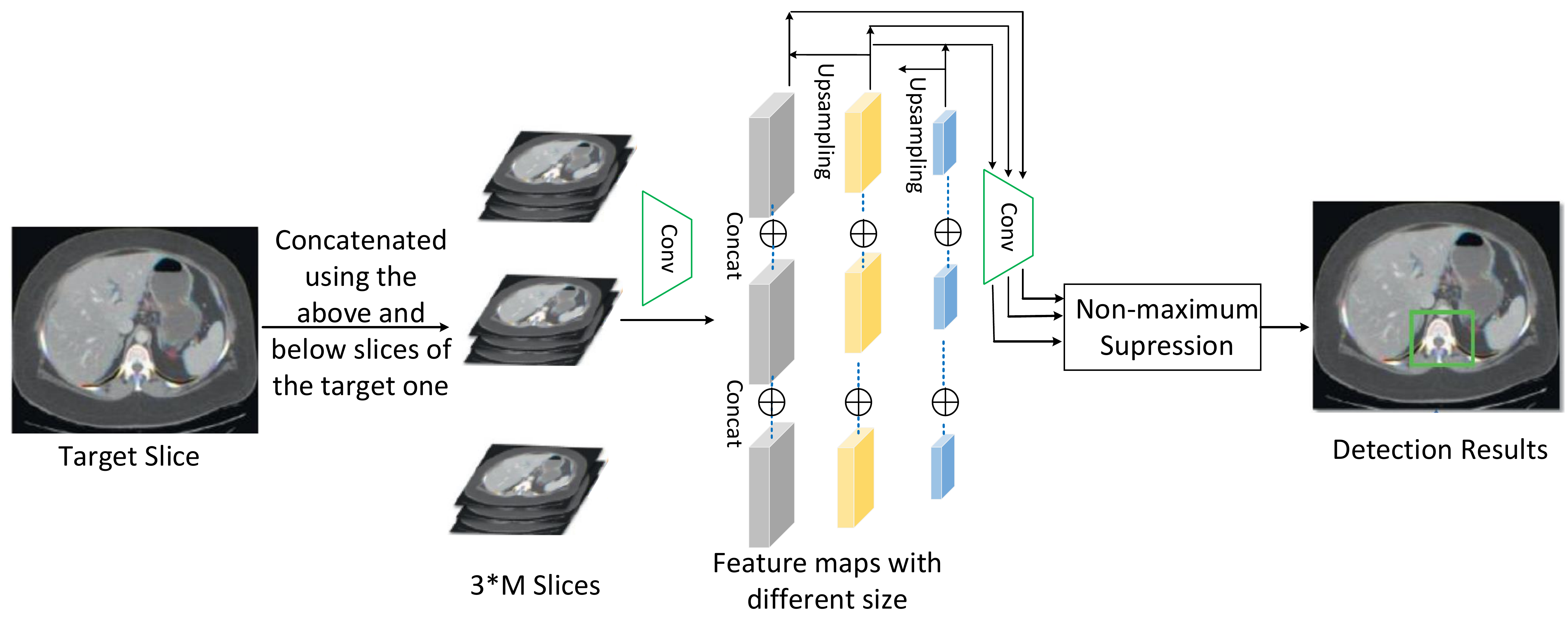}
    \centering
    \caption{
    An example of using one-stage structure for lesion detection in CT images \cite{cai2019one}.
    }
    \label{fig:det_structure31}
    \end{centering}
  \end{figure}

Another category is one-stage models like YOLO (You Only Look Once) \cite{redmon2016you}, SSD (Single Shot MultiBox Detector) \cite{liu2016ssd} and RetinaNet \cite{lin2017focal}. These networks skip the region proposal stage and run detection directly by considering the probability that the object appears at each point in image. Compared with the two-stage models, models in this approach are generally faster and simpler. Examples in this category can be found in breast tumor detection in mammograms \cite{platania2017automated},  pulmonary lung nodule detection in CT \cite{li2017detection}, and the detection of different lesions (e.g., liver lesion, lung lesion, bone lesion, abdomen lesion) in CT images \cite{cai2019one}. An example of using one-stage structure for lesion detection is shown in Fig. \ref{fig:det_structure31}.

In the following subsections, we will introduce related works that incorporate external knowledge into deep learning models for object detection in medical images. A summary of these works is shown in Table \ref{tab:detection_knowledge}. 

\begin{table*}[]
\small
\center
\caption{List of studies on lesion, organ, and abnormality detection and the knowledge they incorporated}
\begin{tabular}{|c|l|l|l|}
\hline
\begin{tabular}[c]{@{}l@{}}Knowledge Source\end{tabular}                 & \begin{tabular}[c]{@{}l@{}}Knowledge Type\end{tabular}                                                                                                             & \begin{tabular}[c]{@{}l@{}}Incorporating Method\end{tabular}                                                                                                                                                & References                                                                                                                                                                                                                                                                                                                                                                                                                                                   \\ \hline
\begin{tabular}[c]{@{}l@{}}Natural datasets\end{tabular}                   & \begin{tabular}[c]{@{}l@{}}Natural images\end{tabular}                                                                                                             & \begin{tabular}[c]{@{}l@{}}Transfer learning\end{tabular}                                                                                                                                                   & 
{\begin{tabular}[c]{@{}l@{}}\cite{shin2016deep}\cite{yap2018automated}
\cite{nappi2016deep}\cite{zhang2016automatic} \cite{tajbakhsh2016convolutional} \\ \end{tabular}} \\ \hline
\begin{tabular}[c]{@{}l@{}}Medical datasets\end{tabular} & \begin{tabular}[c]{@{}l@{}}Multi-modal images\end{tabular}                                                                                                         & \begin{tabular}[c]{@{}l@{}}Transfer learning\end{tabular}                                                                                     & \begin{tabular}[c]{@{}l@{}}\cite{zhang2018breast}\cite{ben2019cross}\cite{zhao2020tripartite}\end{tabular}                                                                                                                                                                                                                                                                                                                                                                                                                                                             \\ \hline
\multirow{8}{*}{\begin{tabular}[c]{@{}l@{}}Medical doctors\end{tabular}} & \begin{tabular}[c]{@{}l@{}}Training pattern\end{tabular}                                                                                                           & \begin{tabular}[c]{@{}l@{}}Curriculum learning\end{tabular}                                                                                                                                                 &\begin{tabular}[c]{@{}l@{}}\cite{tang2018attention}\cite{jesson2017cased}
\cite{astudillo2020curriculum}\end{tabular}                                                                                                                                                                                                                                                                                                                                                                                                                                                              \\ \cline{2-4}
                                 & \multirow{5}{*}{\begin{tabular}[c]{@{}l@{}}Detection patterns\end{tabular}}                                                                                                         & \begin{tabular}[c]{@{}l@{}}Using images collected \\under different settings \end{tabular}                                                                                                                                       & \cite{li2019mvp-net}\cite{ni2020deep}                                                                                                                                                                                                                                                                                                                                                                                                                                                              \\ \cline{3-4}
                                 & & \begin{tabular}[c]{@{}l@{}}Comparing bilateral or \\cross-view  images \end{tabular}& \begin{tabular}[c]{@{}l@{}}\cite{liu2019from}\cite{2020Cross}  \cite{lisowska2017thrombus,lisowska2017context}
                                 \cite{li2020deep}\end{tabular}
                                 \\ \cline{3-4}
                                 & & \begin{tabular}[c]{@{}l@{}}Analyzing adjacent slices\end{tabular} & \cite{li2017detection}
                                 \\ \cline{2-4}
                                 & \begin{tabular}[c]{@{}l@{}}Features doctors 
                                 focus on\end{tabular}                                      & \begin{tabular}[c]{@{}l@{}}Feature level fusion\end{tabular}&
                                 \begin{tabular}[c]{@{}l@{}}\cite{fu2017automatic}\cite{kooi2017large} \cite{ghatwary2019esophageal}\cite{liu2019automated}
                                 \cite{chao2020lymph}\cite{sonora2020evaluating}\end{tabular}
                                  \\ \cline{2-4}
                                 & \begin{tabular}[c]{@{}l@{}}Other related information 
                                 \end{tabular} & \begin{tabular}[c]{@{}l@{}}Multi-task  learning \\/training pattern design\end{tabular}                                                                                                                                                           &\begin{tabular}[c]{@{}l@{}}\cite{tang2018attention}\cite{hwang2016self}
                                 \cite{bakalo2019classification}\cite{liang2020weakly}\end{tabular}                                                                                                                                                                                                                                                                                                                                                                                                                                                             \\ \hline
\end{tabular}
\label{tab:detection_knowledge}
\end{table*}


\subsection{Incorporating Knowledge from Natural Datasets or Other Medical Datasets}
\label{sec:det_natural}

Similar to disease diagnosis, it is quite popular to pre-train a large natural image dataset (generally ImageNet) to introduce information for object detection in medical domain. Examples can be found in lymph node detection \cite{shin2016deep}, polyp and pulmonary embolism detection \cite{tajbakhsh2016convolutional}, breast tumor detection \cite{yap2018automated}, colorectal polyps detection \cite{nappi2016deep,zhang2016automatic} and so on.

In addition, utilizing other medical datasets (i.e., multi-modal images) is also quite common. For example,  PET images are used to help the lesion detection in CT scans of livers \cite{ben2019cross}. Specifically, PET images are first generated from CT scans using a combined structure of FCN and GAN, then the synthesized PET images are used in a false-positive reduction layer for detecting malignant lesions. Quantitative results show a 28\% reduction in the average false positive per case. Besides, \cite{zhang2018breast} develops a strategy to detect breast masses from digital tomosynthesis by fine-tuning the model pre-trained on mammography datasets. Another example of using multi-modal medical images can be found in liver tumor detection \cite{zhao2020tripartite}.

\subsection{Incorporating Knowledge from Medical Doctors}
\label{sec:det_radiologists}

In this subsection, we summarize the existing works on incorporating the  knowledge of medical doctors to more effectively detect objects in medical images. The types of domain knowledge from medical doctors can be grouped into the following four categories:
\begin{enumerate}
\item the training pattern,
\item the general detection patterns they view images,
\item the features (e.g., locations, structures, shapes) they give special attention to, and
\item other related information regarding detection.
\end{enumerate}


\subsubsection{Training Pattern of Medical Doctors}
\label{sec:det_trainingpattern}

The training pattern of medical doctors, which is generally characterized as being  given tasks with increasing difficulty, is also used for object detection in medical images. Similar to the disease diagnosis, most works utilize the curriculum learning to introduce this pattern. For example, an attention-guided curriculum learning (AGCL) framework is presented to locate the lesion in chest X-rays \cite{tang2018attention}. During this process, images in order of difficulty (grouped by different severity-levels) are fed into the CNN gradually, and the disease heatmaps generated from the CNN are used to locate the lesion areas.

Another work is called as CASED proposed for lung nodule detection in chest CT \cite{jesson2017cased}. CASED adopts a curriculum adaptive sampling technique to address the problem of extreme data imbalance. In particular, CASED lets the network to first learn how to distinguish nodules from their immediate surroundings, and then it adds a greater proportion of difficult-to-classify global context, until uniformly samples from the empirical data distribution. In this way, CASED tops the LUNA16 challenge leader-board with an average sensitivity score of 88.35\%. Another example of using curriculum learning can be found in cardiac landmark detection \cite{astudillo2020curriculum}.

\subsubsection{General Detection Patterns of Medical Doctors}
\label{sec:det_diagnosticpattern}

When experienced medical doctors are locating possible lesions in medical images, they also display particular patterns, and these patterns can be incorporated into deep learning models for object detection.
Experienced medical doctors generally have the following patterns: (1) they usually take into account images collected under different settings (e.g., brightness and contrast), (2) they often compare bilateral images or use cross-view images, and (3) they generally read adjacent slices.


For example, to locate possible lesions during visual inspection of the CT images, radiologists would combine images collected under different settings (e.g., brightness and contrast). To imitate the above process, a multi-view feature pyramid network (FPN) is proposed in \cite{li2019mvp-net}, where multi-view features are extracted from images rendered with varied brightness and contrast. The multi-view information is then combined using a position-aware attention module. Experiments show that the proposed model achieves an absolute gain of 5.65\% over the previous state-of-the-art method
on the NIH DeepLesion dataset. Another example of using images under different settings can be found in the COVID-19 pneumonia-based lung lesion detection \cite{ni2020deep}.



\begin{figure}[!htp] \begin{centering}
   \includegraphics[width=0.75\linewidth]{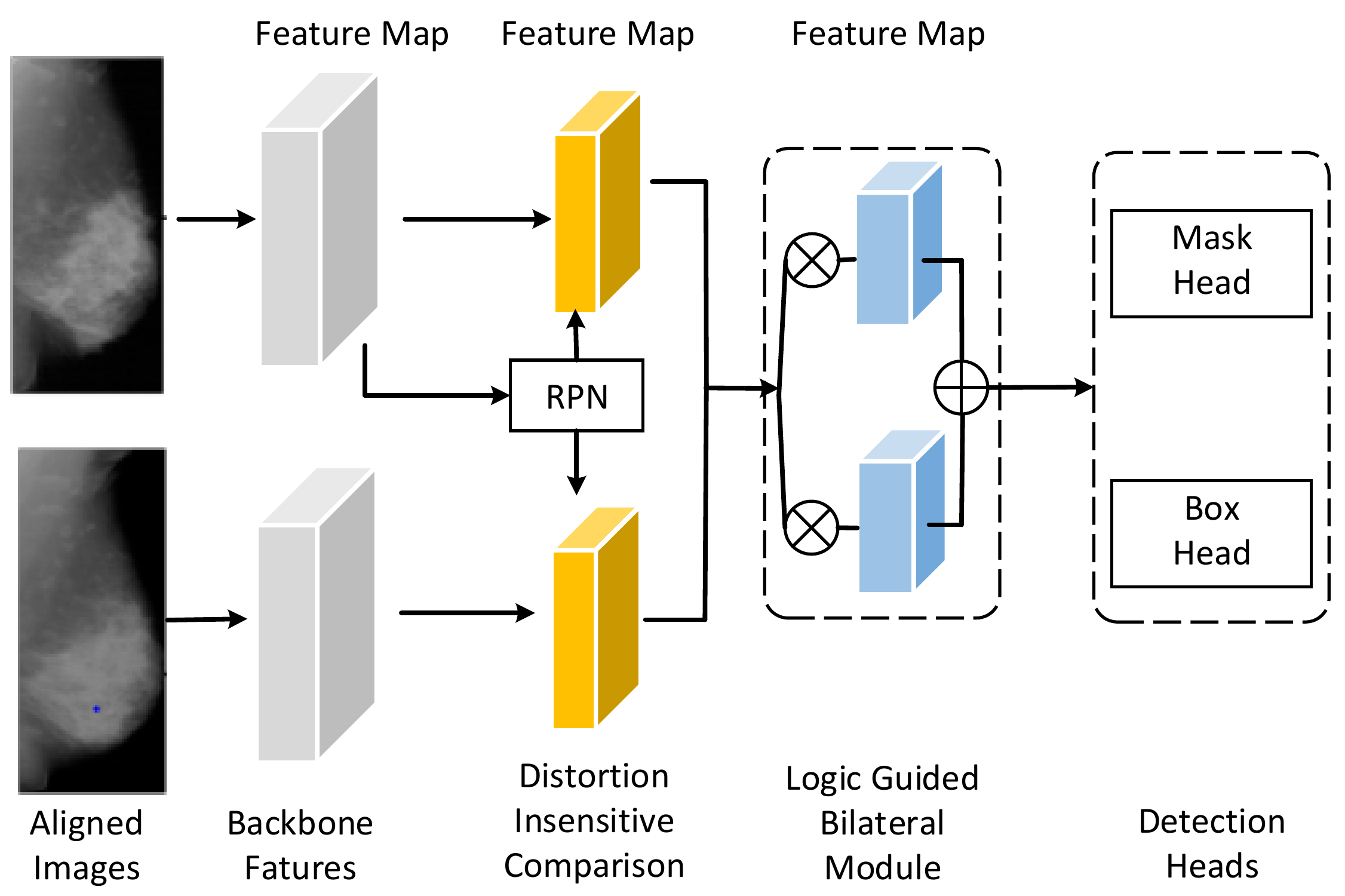}
   \centering
   \caption{The workflow of mammogram mass detection by integrating the bilateral information \cite{liu2019from}, where the aligned images are fed into two networks seperately to extract features for further detection.}
   \label{fig:bilateral_image}
   \end{centering}
 \end{figure}

In addition, the bilateral information is widely used by radiologists. For example, in standard mammographic screening, images are captured from both two breasts, and experienced radiologists generally compare bilateral mammogram images to find masses. To incorporate this pattern, a contrasted bilateral network (CBN) is proposed in \cite{liu2019from}, in which the bilateral images are first coarsely aligned and then fed into a pair of networks to extract features for the following detection steps (shown in Fig. \ref{fig:bilateral_image}). Experimental results demonstrate the effectiveness of incorporating the bilateral information.


Similarly, to detect acute stroke signs in non-contrast CT images, experienced neuroradiologists routinely compare the appearance and Hounsfield Unit intensities of the left and right hemispheres, and then find the regions most commonly affected in stroke episodes. This pattern is mimicked by \cite{lisowska2017context} for the detection of dense vessels and ischaemia. The experimental results show that introducing the pattern greatly improves the performance for detecting ischaemia. Other examples of integrating the bilateral feature comparison or the symmetry constrains can be found in thrombus detection \cite{lisowska2017thrombus} and  hemorrhagic lesion detection \cite{li2020deep} in brain CT images.

Besides the bilateral images, the information from cross views (i.e., mediolateral-oblique and cranio-caudal) is highly related and complementary, and hence is also used for mammogram mass detection. In \cite{2020Cross}, a bipartite graph convolutional network is introduced to endow the existing methods with cross-view reasoning ability of radiologists. Concretely, the bipartite node sets are constructed to represent the relatively consistent regions, and the bipartite edge are used to model both inherent cross-view geometric constraints and appearance similarities between correspondences. This process can enables spatial visual features equipped with cross-view reasoning ability. Experimental results on DDSM dataset achieve the state-of-the-art performance (with a recall of 79.5 at 0.5 false positives per image). 


When looking for small nodules in CT images , radiologists often observe each  slice together with adjacent slices, similar to detecting an object in a video. This workflow is imitated in \cite{li2017detection} to detect pulmonary lung nodule in CT images, where the state-of-the-art object detector SSD is applied in this process. This method obtains state-of-the-art result with the FROC score of 0.892 in LUNA16 dataset.

\subsubsection{Features That Medical Doctors Give Special Attention to}
\label{sec:det_feature}

Similar to disease diagnosis, medical doctors also use many `hand-crafted  features' to help them to find target objects (e.g., nodules or lesions) in medical images. 


For example, in \cite{kooi2017large}, to detect mammographic lesions, different  types of hand-crafted features (e.g., contrast features, geometrical features, and location features) are first extracted, and then concatenated with those learned by  a CNN. 
The results show that these hand-crafted features, particularly the location and context features 
, can complement the network generating a higher specificity over the CNN alone.


Similarly, \cite{ghatwary2019esophageal} presents a deep learning model based on Faster R-CNN to detect abnormalities in the esophagus from endoscopic images. In particular, to enhance texture details, the proposed detection system incorporates the Gabor handcrafted features
with the CNN features through concatenation in the detection stage.
The experimental results on two datasets (Kvasir and MICCAI 2015) show that the model is able to surpass the state-of-the-art performance.

Another example can be found in  \cite{fu2017automatic} for the detection of lung nodules, where 88 hand-crafted features, including intensity, shape, texture are extracted and combined with features extracted by a CNN and then fed into a classifier. Experimental results
demonstrate the effectiveness of  the combination of handcrafted features and
CNN features.

In the automated detection of thyroid nodules, the size and shape attribute of nodules are considered in \cite{liu2019automated}. To incorporate the information above, the generating process of region proposals is constrained, and the detection results on two different datasets show high accuracy.

Furthermore, in lymph node gross tumor volume detection (GTV$_{LN}$) in oncology imaging, some attributes of lymph nodes (LNs) are also utilized \cite{chao2020lymph}. Motivated by the prior clinical knowledge that LNs from a connected lymphatic system, and the spread of cancer cells among LNs often follows certain pathways, a LN appearance and inter-LN relationship learning framework is proposed for GTV$_{LN}$ detection. More specifically, the instance-wise appearance features are first extracted by a 3D CNN, then a graph neural network (GNN) is used to model the inter-LN relationships, and the global LN-tumor spatial priors are included in this process. This method
significantly improves over state-of-the-art method. Another example of combining handcrafted features and deep features can be found in lung lesion detection \cite{sonora2020evaluating}.

\subsubsection{Other Types of Information Related to Detection}
\label{sec:det_extraannotations}
Similar with that in disease diagnosis, there are other information (e.g., radiological reports, extra labels) can also be integrated into the lesion detection process.

For example in \cite{tang2018attention}, to locate thoracic diseases on chest radiographs, the difficulty of each sample, represented as the severity level of the thoracic disease, is first extracted from radiology reports. Then, the curriculum learning technique is adopted, in which the training samples are presented to the network in order of increasing difficulty. Experiments on the ChestXray14 database validate the effectiveness on significant performance improvement over baseline methods.

Example of using extra labels can be found in \cite{hwang2016self}. In this method, the information of the classification labels is incorporated to help the lesion localization in chest X-rays and mammograms. In particular, a framework named as self-transfer learning (STL) is proposed, which jointly optimizes both classification and localization networks to help the localization network focus on correct lesions. Experimental results show that STL can achieve significantly better localization performance compared to previous weakly supervised localization approaches. More examples of using extra labels can be found in detection in mammograms \cite{bakalo2019classification,liang2020weakly}.

\subsection{Summary}
\label{sec:det_overview}

In the previous sections, we introduced different kinds of domain knowledge and the corresponding integrating methods into the deep learning models for object detection in medical images.  Table  \ref{tab:detquan} illustrates the quantitative improvements, in terms of sensitivity and recall, of some typical work over the baseline methods for object detection in medical images. From the results we can see that in general, integrating domain knowledge can be beneficial for detection tasks.

Similar to disease diagnosis, the high-level training pattern of medical doctors is generic and can be utilized for detecting different diseases or organs. In contrast, the low-level domain knowledge, like the detection patterns that medical doctors adopt and some hand-crafted features they give more attention when searching lesions, are generally different for different diseases. For example, the pattern of comparing bilateral images can only be utilized for detecting organs with symmetrical structures \cite{lisowska2017context,liu2019from}. In addition, we can see from Table \ref{tab:detquan} that leveraging pattern of medical doctors on average shows better performance when compared with integrating hand-crafted features. This may indicate that there is still large room to explore more effective features for object detection in medical images.

\begin{table*}[]
\caption{The comparison of the quantitative metrics for some medical object detection methods after incorporating domain knowledge}
\small
\center
\begin{threeparttable}
\begin{tabular}{|l|l|l|l|}
\cline{1-4}
\multicolumn{1}{|c|}{Reference}                              & \multicolumn{1}{c|}{\begin{tabular}[c]{@{}l@{}}Baseline model/With domain knowledge\end{tabular}}                              & \multicolumn{1}{c|}{Sensitivity}   & \multicolumn{1}{c|}{Recall}           \\ \hline
\multirow{1}{*}{\cite{fu2017automatic}} & \multicolumn{1}{c|}{\begin{tabular}[c]{@{}l@{}}CNN/CNN with hand-crafted features\end{tabular}} & \multicolumn{1}{c|}{0.890/0.909} &\multicolumn{1}{c|}{--/--}    \\ \hline
\multirow{1}{*}{\cite{ghatwary2019esophageal}} & \multicolumn{1}{c|}{\begin{tabular}[c]{@{}l@{}}ResNet/CNN with handcrafted features\end{tabular}} & \multicolumn{1}{c|}{--/--} &\multicolumn{1}{c|}{0.940/0.950}    \\ \hline
\multirow{1}{*}{\cite{li2017detection}} & \multicolumn{1}{c|}{SSD/MSSD} & \multicolumn{1}{c|}{0.927/0.976}&\multicolumn{1}{c|}{--/--}  \\ \hline
\multirow{1}{*}{\cite{liu2019from}} & \multicolumn{1}{c|}{Mask R-CNN/CBD}       & \multicolumn{1}{c|}{--/--} &\multicolumn{1}{c|}{0.869/0.890}  \\ \hline
\multirow{1}{*}{\cite{2020Cross}} & \multicolumn{1}{c|}{Mask R-CNN/BG-RCNN}        &  \multicolumn{1}{c|}{--/--} &\multicolumn{1}{c|}{0.918/0.945}  \\ \hline
\multirow{1}{*}{\cite{tang2018attention}} & \multicolumn{1}{c|}{\begin{tabular}[c]{@{}l@{}}CNN/AGCL\end{tabular}}&  \multicolumn{1}{c|}{--/--} &\multicolumn{1}{c|}{0.660/0.730}  \\ \hline
\end{tabular}
\label{tab:detquan}
\begin{tablenotes}
\footnotesize
\item[1] The performance is evaluated at 4 false positives per image in \cite{fu2017automatic,liu2019from,2020Cross,samala2018breast}.
\end{tablenotes}
\end{threeparttable}
\end{table*}
\section{Lesion and Organ Segmentation}
\label{sec:segmentation}

Medical image segmentation devotes to identifying pixels of lesions or organs from the background, and is generally regarded as a prerequisite step for the lesion assessment and disease diagnosis. Segmentation methods based on deep learning models have become the dominant technique in recent years and have been widely used for the segmentation of lesions such as brain tumors \cite{havaei2017brain}, breast tumors \cite{zhang2018hierarchical}, and  organs such as livers \cite{christ2017automatic} and pancreas \cite{roth2015deep}.

In Subsection \ref{sec:seg_structures}, we describe the models that are generally used for object segmentation in the medical domain. Then in Subsection \ref{sec:segmentation_natural}, the works of utilizing domain knowledge from natural and other medical datasets are introduced. Then, the models utilizing domain knowledge from medical doctors are introduced in Subsection \ref{sec:seg_radiologists}. A summary of this section is provided in Subsection \ref{sec:seg_overview}.

\subsection{General Structures of Deep Learning Models for Object Segmentation in Medical Images}
\label{sec:seg_structures}

The deep learning models utilized for medical image segmentation are generally divided into three categories: the FCN (fully convolutional network) \cite{long2015fully} based models, the U-Net \cite{ronneberger2015u} based models, and the GAN \cite{goodfellow2014generative} based models.

\begin{figure}[!htp] \begin{centering}
   \includegraphics[width=0.85\linewidth]{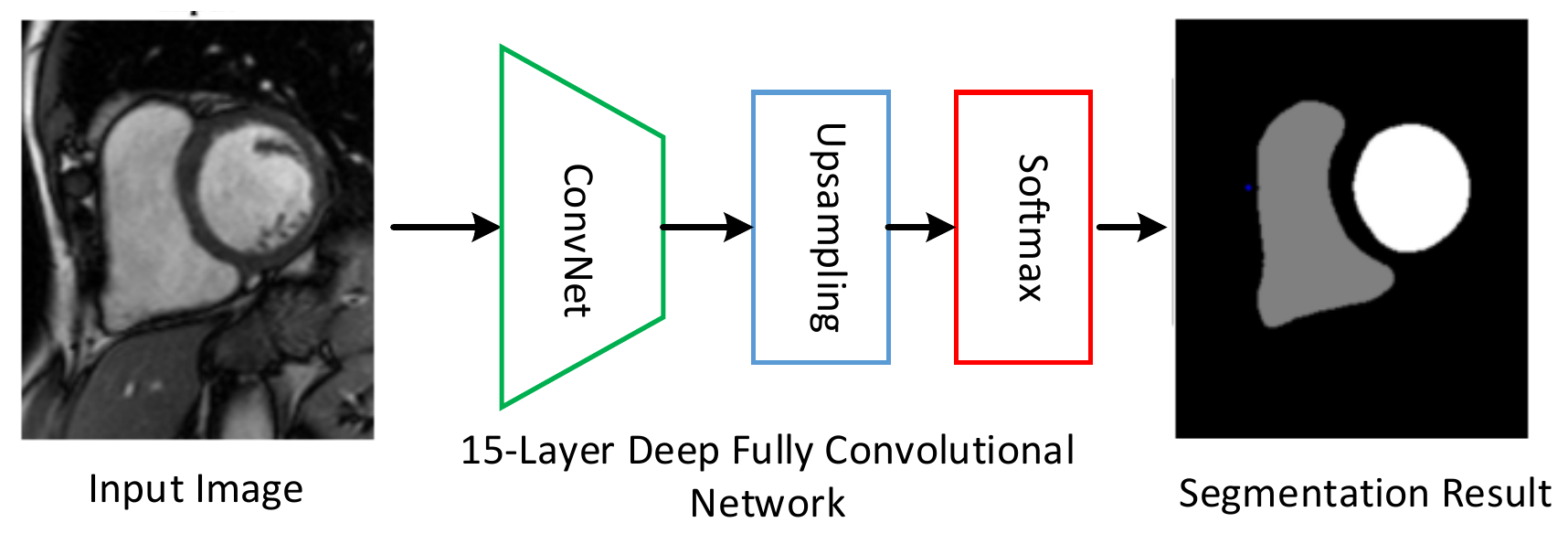}
   \centering
   \caption{The schematic diagram of using FCN structure for cardiac segmentation  \cite{tran2016fully}.}
   \label{fig:seg_FCN}
   \end{centering}
 \end{figure}
In particular, the FCN has been proven to perform well in various medical image segmentation tasks \cite{chen2016iterative,gibson2017towards}. Some  variants of FCN, such as cascaded FCN \cite{christ2016automatic}, parallel FCN \cite{kamnitsas2017efficient} and recurrent FCN \cite{yang2017fine} are also widely used for segmentation tasks in medical images. Fig. \ref{fig:seg_FCN} illustrates an example of using FCN based model for cardiac segmentation.

\begin{figure}[!htb] \begin{centering}
   \includegraphics[width=0.9\linewidth]{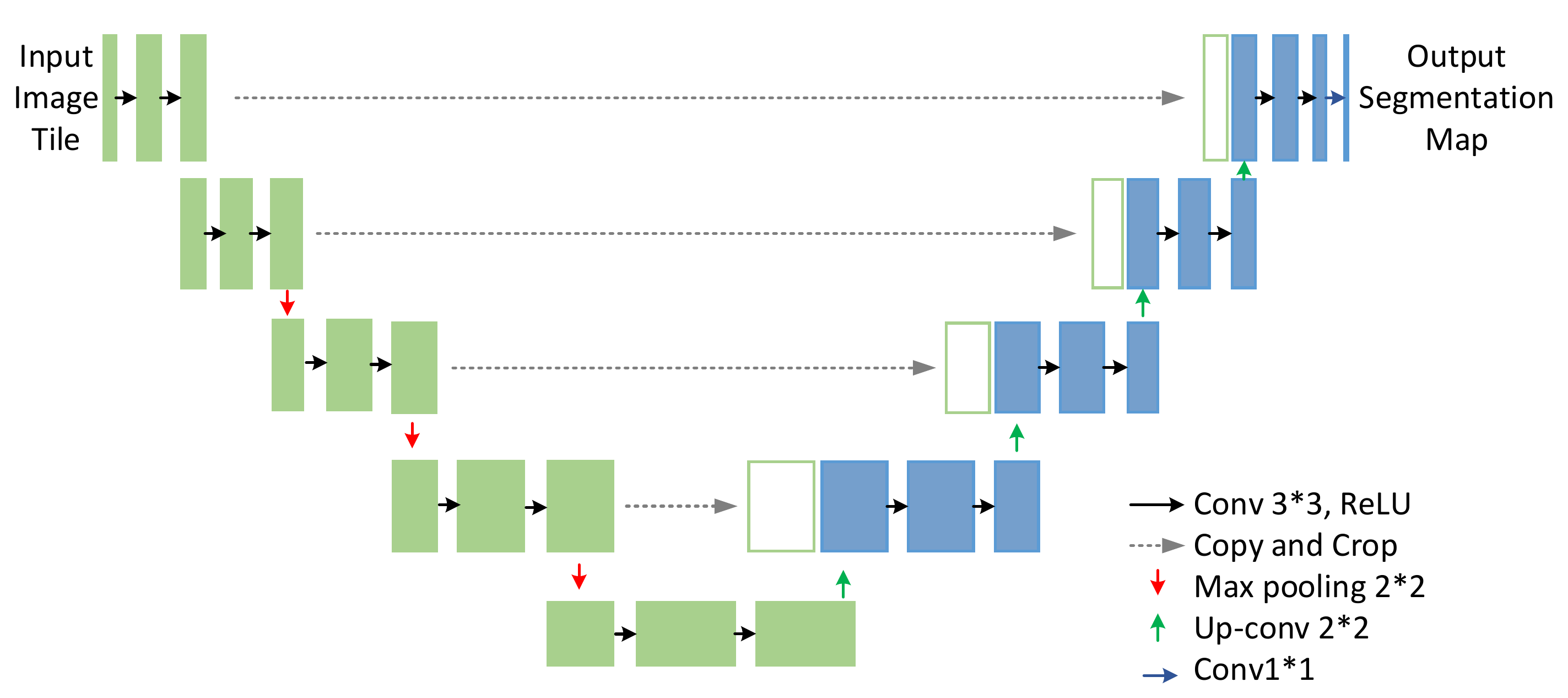}
   \centering
   \caption{The network structure of U-Net \cite{ronneberger2015u}.
   }
   \label{fig:seg_structure_UNet}
   \end{centering}
\end{figure}

In addition, the U-Net \cite{ronneberger2015u} (shown in Fig. \ref{fig:seg_structure_UNet}) and its variants are also widely utilized for medical image segmentation. U-Net builds upon FCN structure, mainly consists of a series of convolutional and deconvolutional layers, and with the short connections between the layers of equal resolution. U-Net and its variants like UNet++\cite{zhou2018unet++} and recurrent U-Net \cite{alom2018recurrent} perform well in many medical image segmentation tasks
\cite{gordienko2018deep}. 

\begin{figure}[!htp] \begin{centering}
   \includegraphics[width=0.85\linewidth]{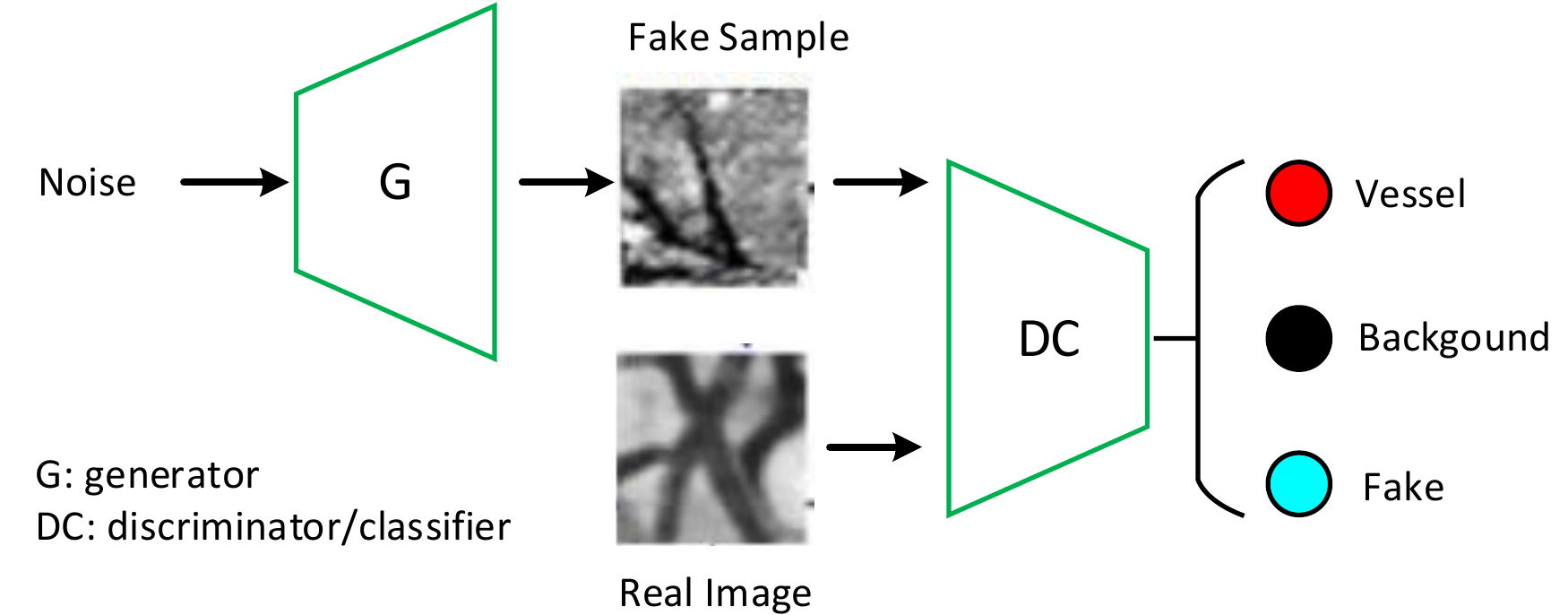}
   \centering
   \caption{An example of using a GAN-based model for vessel segmentation \cite{lahiri2017generative}.}
   \label{fig:seg_GAN}
   \end{centering}
 \end{figure}

In the GAN-based models \cite{yang2017automatic,zhao2018craniomaxillofacial}, the generator is used to predict the mask of the target based on some encoder-decoder structures (such as FCN or U-Net).  
The discriminator serves as a shape regulator that helps the generator to obtain satisfactory segmentation results. Applications of using GAN-based models in medical image segmentation include brain segmentation \cite{kamnitsas2017unsupervised}, skin lesion segmentation \cite{izadi2018generative}, vessel segmentation \cite{lahiri2017generative} and anomaly segmentation in retinal fundus images \cite{schlegl2017unsupervised}. Fig. \ref{fig:seg_GAN} is an example of using GAN-based model for vessel segmentation in miscroscopy images.

\begin{table*}[]
\small
\center
\caption{The list of researches of lesion, organ segmentation and the knowledge they incorporated}
\begin{tabular}{|c|l|l|l|}
\hline
\begin{tabular}[c]{@{}l@{}}Knowledge Source\end{tabular}                 & \begin{tabular}[c]{@{}l@{}}Knowledge Type\end{tabular}                                                                                                             & \begin{tabular}[c]{@{}l@{}}Incorporating\\ Method\end{tabular}                                                                                                                                                & References                                                                                                                                                                                                                                                                                                                                                                                                                                                   \\ \hline
\begin{tabular}[c]{@{}l@{}}Natural datasets\end{tabular}                   & \begin{tabular}[c]{@{}l@{}}Natural images\end{tabular}                                                                                                             & \begin{tabular}[c]{@{}l@{}}Transfer learning\end{tabular}                                                                                                                                                   & 
{\begin{tabular}[c]{@{}l@{}} \cite{chen2016iterative}\cite{chen2017dcan}\cite{wu2017cascaded}\cite{zeng20173d}
\cite{tajbakhsh2016convolutional}
\end{tabular}} \\ \hline
\multirow{6}{*}{\begin{tabular}[c]{@{}l@{}}Medical datasets\end{tabular}} & \multirow{4}{*}{\begin{tabular}[c]{@{}l@{}}Multi-modal images\end{tabular}}                                                                                                         & \begin{tabular}[c]{@{}l@{}}Transfer learning\end{tabular}                                                                                     & \begin{tabular}[c]{@{}l@{}}\cite{ghafoorian2017transfer}\cite{valverde2019one-shot}\end{tabular}                                                                                                                                                                                                                                                                                                                                                                                                                                                             \\ \cline{3-4}
& & \begin{tabular}[c]{@{}l@{}}Multi-task/-modal learning\end{tabular}                                                                                      & 
{\begin{tabular}[c]{@{}l@{}}\cite{moeskops2016deep}\cite{valindria2018multi}\end{tabular}}                                                                                                                                                                                                                                                                                                                                                                                                                                                             \\ \cline{3-4}
& & \begin{tabular}[c]{@{}l@{}}Modality transformation  /synthesis
\end{tabular}                                                                                     & 
{\begin{tabular}[c]{@{}l@{}}\cite{chen2018semantic}\cite{chen2020unsupervised} \cite{jiang2018tumor} \cite{kamnitsas2017unsupervised} \cite{yan2019domain}\\\cite{yang2019unsupervised}{\cite{li2020towards}\cite{li2020dual}\cite{hu2020knowledge}}
\cite{zhang2018translating}\cite{chartsias2020disentangle}
\end{tabular}}                                                                                                                                                                                                                                                                                                                                                                                                                                                            \\ \cline{2-4}
                                 & \multirow{2}{*}{\begin{tabular}[c]{@{}l@{}}Datasets of other diseases\end{tabular}}                                                                                               & Transfer learning                                                                                                                                                 & \cite{chen2019med3d}                                                                                                                                                                                                                                                                                                                                                                                                                                                            \\ \cline{3-4}
                                 & & \begin{tabular}[c]{@{}l@{}}Disease domain  transformation 
                                 \end{tabular}
                                 &
                                 {\cite{yu2019annotation}}
                                 \\ \hline
\multirow{11}{*}{\begin{tabular}[c]{@{}l@{}}Medical doctors\end{tabular}} & \begin{tabular}[c]{@{}l@{}}Training pattern\end{tabular}                                                                                                           & \begin{tabular}[c]{@{}l@{}}Curriculum learning\end{tabular}                                                                                                                                                 &\begin{tabular}[c]{@{}l@{}}\cite{berger2018adaptive} \cite{wang2018deep}
\cite{li2020new}\cite{kervadec2019curriculum}
 \cite{zhao2019semi}\cite{zhang2020weakly}\end{tabular}                                                                                                                                                                                                                                                                                                                                                                                                                                                              \\ \cline{2-4}
                                 & \multirow{3}{*}{\begin{tabular}[c]{@{}l@{}}Diagnostic  patterns\end{tabular}}                                                                                                         & \begin{tabular}[c]{@{}l@{}}Using different views as input\end{tabular}                                                                                                                                       & \begin{tabular}[c]{@{}l@{}}\cite{wu2018joint}\cite{chen2019learning}\end{tabular}                                                                                                                                                                                                                                                                                                                                                                                                                                                              \\ \cline{3-4}
                                 & & \begin{tabular}[c]{@{}l@{}}Attention mechanism\end{tabular} & \cite{hatamizadeh2019end}
                                 \\ \cline{3-4}
                                 & & \begin{tabular}[c]{@{}l@{}}Network design\end{tabular} & \cite{zhu2020lymph}\cite{jin2020deeptarget}
                                 \\ \cline{2-4}
                                 & \multirow{5}{*}{\begin{tabular}[c]{@{}l@{}}Anatomical priors\end{tabular}}                                      & \begin{tabular}[c]{@{}l@{}}Incorporated in post-processing 
                                 \end{tabular} &
                                 \begin{tabular}[c]{@{}l@{}}\cite{huang2018medical} \cite{painchaud2019cardiac,painchaud2020cardiac}
                                 \cite{larrazabal2019anatomical}\end{tabular}
                                 \\ \cline{3-4}
                                  & & \begin{tabular}[c]{@{}l@{}}Incorporated in loss function\end{tabular} &
                                 \begin{tabular}[c]{@{}l@{}}\cite{oktay2017anatomically}\cite{ravishankar2017learning} \cite{bentaieb2016topology}\cite{mirikharaji2018star}  \cite{yue2019cardiac}\\\cite{zheng2019semi-supervised}
                                 \cite{oktay2017anatomically}                        \cite{zotti2019convolutional}
                                 \end{tabular}
                                  \\ \cline{3-4}
                                  & & \begin{tabular}[c]{@{}l@{}}Generative models \end{tabular}&
                                 \begin{tabular}[c]{@{}l@{}}\cite{chen2019learning}\cite{dalca2018anatomical} \cite{he2019dpa-densebiasnet}\cite{luo2020shape}
                                 \cite{song2020shape}\\\cite{boutillon2020combining}
                                  \cite{pham2020liver}\cite{engin2020agan}
                                  \cite{gao2020focusnetv2}\end{tabular}
                                  \\ \cline{2-4}
                                 & \multirow{2}{*}{\begin{tabular}[c]{@{}l@{}}Hand-crafted  features\end{tabular}} & \begin{tabular}[c]{@{}l@{}}Feature level fusion\end{tabular}                                                                                                                                                           &\begin{tabular}[c]{@{}l@{}}\cite{kushibar2018automated}\cite{rezaei2019gland}\end{tabular}
                                 \\ \cline{3-4}
                                 & & \begin{tabular}[c]{@{}l@{}}Input level fusion\end{tabular}                                                                                                                                                           &\begin{tabular}[c]{@{}l@{}}\cite{khan2020cascading}\cite{narotamo2020combining}  \end{tabular}                                                                                                                                                                                                                                                                                                                                                                                                                                                           \\ \hline
\end{tabular}
\label{tab:segmentation_knowledge}
\end{table*}
In the following sections, we will introduce research studies that incorporate domain knowledge into deep segmentation models. The summary of these works is shown in Table \ref{tab:segmentation_knowledge}.

\subsection{Incorporating Knowledge from Natural Datasets or Other Medical Datasets}
\label{sec:segmentation_natural}

It is also quite common that deep learning segmentation models are firstly trained on a large-scale natural image dataset (e.g., ImageNet) and then fine-tuned on the target datasets. Using the above transfer learning strategy to introduce knowledge from natural images has demonstrated to achieve a better performance in medical image segmentation. Examples can be found in intima-media boundary segmentation \cite{tajbakhsh2016convolutional} and prenatal ultrasound image segmentation\cite{wu2017cascaded}. Besides ImageNet, \cite{chen2016iterative} adopts the off-the-shelf DeepLab model trained on the PASCAL VOC dataset for anatomical structure segmentation in ultrasound images. This pre-trained model is also used in the deep contour-aware network (DCAN), which is designed for the gland segmentation in histopathological images \cite{chen2017dcan}.

Besides using models pre-trained on `static' datasets like ImageNet and PASCAL VOC, many deep neural networks, especially those designed for the segmentation of 3D medial images, leverage models pre-trained on large-scale video datasets. For example, in the automatic segmentation of proximal femur in 3D MRI, the C3D pre-trained model is adopted as the encoder of the proposed 3D U-Net \cite{zeng20173d}. Notably, the C3D model is trained on the Sports-1M dataset, which is the largest video classification benchmark with 1.1 million sports videos in 487 categories \cite{tran2015learning}.

In addition to natural images, using knowledge from external medical datasets with different modalities and with different diseases is also quite popular.

For example, \cite{ghafoorian2017transfer} investigates the transferability of the acquired knowledge of a CNN model initially trained for WM hyper-intensity segmentation on legacy low-resolution data to new data from the same scanner but with higher image resolution. Likewise, the images with different MRI scanners and protocols are used in \cite{valverde2019one-shot} to help the multi sclerosis segmentation process via transfer learning. 


In \cite{moeskops2016deep}, the multi-task learning is adopted,
where the data of brain MRI, breast MRI and cardiac CT angiography (CTA) are used
simultaneously as multiple tasks. On the other hand, \cite{valindria2018multi} adopts
a multi-modal learning structure for organ segmentation. A dual-stream encoder-decoder
architecture is proposed to learn modality-independent, and thus, generalisable and
robust features shared among medical datasets with different modalities (MRI and CT images).
Experimental results prove the effectiveness of this multi-modal learning structure.


Moreover, many works adopt GAN-based models to achieve the domain transformation among
datasets with different modalities. For example, a model named SeUDA (unsupervised domain
adaptation) is proposed for the left/right lung segmentation process \cite{chen2018semantic}.
It leverages the semantic-aware GAN to transfer the knowledge from one chest dataset to
another. In particular, target images are first mapped towards the source data space via the
constraint of a semantic-aware GAN loss. Then the segmentation results are obtained
from the segmentation DNN learned on the source domain. Experimental results show that
the segmentation performance of SeUDA is highly competitive.


More examples of using the knowledge from images with other modalities can be found in brain MRI segmentation \cite{kamnitsas2017unsupervised,hu2020knowledge}, cardiac segmentation
\cite{yan2019domain,zhang2018translating,li2020dual,li2020towards,chartsias2020disentangle}, liver segmentation \cite{yang2019unsupervised}, lung tumor segmentation \cite{jiang2018tumor}, 
cardiac substructure and abdominal multi-organ segmentation \cite{chen2020unsupervised}. 

There are also a few works utilize the datasets of other diseases. For instance, \cite{chen2019med3d} first builds a union dataset (3DSeg-8) by aggregating eight different 3D medical segmentation datasets, and designs the Med3D network to co-train based on 3DSeg-8. Then the pre-trained models obtained from Med3D are transferred into lung and liver segmentation tasks. Experiments show that this method not only improves the accuracy, but also accelerates the training convergence speed.
In addition, the annotated retinal images are used to help the cardiac vessel segmentation without annotations \cite{yu2019annotation}. In particular,
a shape-consistent generative adversarial network (SC-GAN) is used to generate the synthetic images and the corresponding labels. Then the synthetic images are used to train the segmentor. Experiments demonstrate the supreme accuracy of coronary artery segmentation. 



\subsection{Incorporating Knowledge from Medical Doctors}
\label{sec:seg_radiologists}

The domain knowledge of medical doctors is also widely adopted when designing deep learning models for segmentation tasks in medical images. The types of domain knowledge from medical doctors utilized in deep segmentation models can be divided into four categories:
\begin{enumerate}
\item the training pattern,
\item the general diagnostic patterns they view images,
\item the anatomical priors (e.g., shape, location, topology) of lesions or organs, and
\item other hand-crafted features they give special attention to.
\end{enumerate}

\subsubsection{Training Pattern of Medical Doctors}
\label{sec:seg_trainingprocess}
Similar to disease diagnosis and lesion/organ detection, many research works for the object segmentation in medical images also mimic the training pattern of medical doctors, which involves assigning tasks that increase in difficulty over time.
In this process, the curriculum learning technique or its derivative methods like self-paced learning (SPL) are also utilized \cite{kumar2010self}.



For example, for the segmentation of multi-organ CT images \cite{berger2018adaptive}, each annotated image is divided into small patches. During the training process, patches producing large error by the network are selected with a high probability. In this manner, the network can focus sampling on difficult regions, resulting in improved performance.
In addition, \cite{wang2018deep} combines the SPL with the active learning for the pulmonary segmentation in 3D images. This system achieves the state-of-the-art segmentation results.

Moreover, a three-stage curriculum learning approach is proposed for liver tumor segmentation \cite{li2020new}. The first stage is performed on the whole input volume to initialize the network, then the second stage of learning focuses on tumor-specific features by training the network on the tumor patches, and finally the network is retrained on the whole input in the third stage. This approach exhibits significant improvement when compared with the commonly used cascade counterpart in MICCAI 2017 liver tumor segmentation (LiTS) challenge dataset. More  examples can also be found in left ventricle segmentation \cite{kervadec2019curriculum}, finger bones segmentation \cite{zhao2019semi} and vessel segmentation \cite{zhang2020weakly}.

\subsubsection{General Diagnostic Patterns of Medical Doctors}
\label{sec:seg_diagnosticpattern}

In the lesion or organ segmentation tasks, some specific patterns that medical doctors adopted are also incorporated into the network.


For example, during the visual inspection of CT images, radiologists often change window widths and window centers to help to make decision on uncertain nodules. This pattern is mimicked in \cite{wu2018joint}. In particular,  image patches of different window widths and window centers are stacked together as the input of the deep learning model to gain rich information. The evaluation implemented on the public LIDC-IDRI dataset indicates that the proposed method achieves promising performance on lung nodule segmentation compared with
the state-of-the-art methods.

In addition, experienced clinicians generally assess the cardiac morphology and function from multiple standard views, using both long-axis (LA) and short-axis (SA) images to form an understanding of the cardiac anatomy. Inspired by the above observation, a cardiac MR segmentation method is proposed which takes three LA and one SA views as the input \cite{chen2019learning}. In particular, the features are firstly extracted using a multi-view autoencoder (MAE) structure, and then are fed in a segmentation network. Experimental results show that this method has a superior segmentation accuracy over
state-of-the-art methods.

Furthermore, expert manual segmentation usually relies on the boundaries of anatomical structures of interest. For instance, radiologists segmenting a liver from CT images would usually trace liver edges first, and then deduce the internal segmentation mask. Correspondingly, boundary-aware CNNs are proposed in  \cite{hatamizadeh2019end} for medical image
segmentation. The networks are designed to account for organ boundary information, both by providing a special network edge branch and edge-aware loss terms. The effectiveness of these boundary aware segmentation networks are tested on BraTS 2018 dataset for the task of brain tumor segmentation.

Recently, the diagnostic pattern named as `divide-and-conquer manner' is mimicked in the GTV$_{LN}$ detection and segmentation method \cite{zhu2020lymph}. Concretely, the GTV$_{LN}$ is first divided into two subgroups of `tumor-proximal' and `tumor-distal', by means of binary of soft distance gating. Then a multi-branch detection-by-segmentation network is trained with each branch specializing on learning one GTV$_{LN}$ category features. After fusing the outs from multi-branch, the method shows significant improvements on the mean recall from 72.5\% to 78.2\%. Another example of using the diagnostic pattern of medical doctors can be found in gross tumor and clinical target volume segmentation \cite{jin2020deeptarget}.

\subsubsection{Anatomical Priors of Lesions or Organs}
\label{sec:seg_feature}

In comparison to non-medical images, medical images have many anatomical priors such as the shape, position and topology of organs or lesions. Experienced medical doctors greatly rely on these anatomical priors when they are doing segmentation tasks on these images. Incorporating the knowledge of anatomical priors into deep learning models has been demonstrated to be an effective way for accurate medical image segmentation. Generally speaking, there are three different approaches to incorporate these anatomical priors into deep learning models: (1) incorporating anatomical priors in the post-processing stage, (2) incorporating anatomical priors as regularization terms in the loss function and (3) learning anatomical priors via generative models.

\textbf{Incorporating anatomical priors in post-processing stage}

The first approach is to incorporate the anatomical priors in the post processing stage. The result of a segmentation network is often blurry and post-processing is generally needed to refine the segmentation result. 

For example, according to the pathology, most of breast tumors begin in glandular tissues and are located inside the mammary layer \cite{sharma2010various}. This position feature is utilized by \cite{huang2018medical} in its post-processing stage where a fully connected conditional random field (CRF) model is employed. In particular, the position of tumors and their relative locations with mammary layer are added as a new term in CRF energy function to obtain better segmentation results.

Besides, some research first learn the anatomical priors, and then incorporate them into the post-processing stage to help produce anatomically plausible segmentation results  \cite{larrazabal2019anatomical,painchaud2019cardiac,painchaud2020cardiac}.
For instance, the latent representation of anatomically correct cardiac shape is first learned by using adversarial variational autoencoder (aVAE), then be used to convert erroneous segmentation maps into anatomically plausible ones \cite{painchaud2019cardiac}. 
Experiments manifest that aVAE is able to accommodate any segmentation method, and convert its anatomically implausible results to plausible ones without affecting its overall geometric and clinical metrics.

\begin{figure}[!htp] \begin{centering}
   \includegraphics[width=0.95\linewidth]{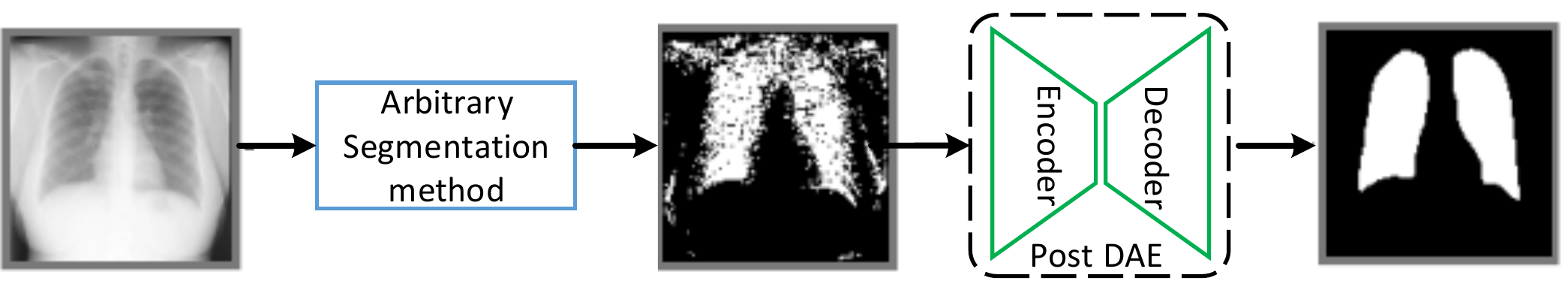}
   \centering
   \caption{The example of integrating the shape prior in the post-process stage  \cite{larrazabal2019anatomical}.
   }
   \label{fig:seg_shape_prior_postprocessing}
   \end{centering}
 \end{figure}

Another example in \cite{larrazabal2019anatomical} introduces the post-processing step based on denoising autoencoders (DAE) for lung segmentation. In particular, the DAE is trained using only segmentation masks, then the learned  representations of anatomical shape and topological constraints are imposed on the original segmentation results (as shown in Fig. \ref{fig:seg_shape_prior_postprocessing}). By applying the Post-DAE on the resulting masks from arbitrary segmentation methods, the lung anatomical segmentation of X-ray images shows plausible results.

\textbf{Incorporating anatomical priors as regularization terms in the loss function}

The second approach is incorporating anatomical priors as regularization terms in the objective function of segmentation networks. For example, for the segmentation of cardiac MR images, a network called as SRSCN is proposed \cite{yue2019cardiac}. SRSCN comprises a shape reconstruction neural network (SRNN) and a spatial constraint network (SCN). SRNN aims to maintain a realistic shape of the resulting segmentation and the SCN is adopted to incorporate the spatial information of the 2D slices.  The loss of the SRSCN comes from three parts: the segmentation loss, the shape reconstruction (SR) loss for shape regularization, and the spatial constraint (SC) loss to assist segmentation. The results using images from 45 patients demonstrate the effectiveness of the SR and SC regularization terms, and show the superiority of segmentation performance of the SRSCN over the conventional schemes.

Another example in this category is the one designed for skin lesion segmentation \cite{mirikharaji2018star}. In this work, the star shape prior is encoded as a new loss term in a FCN to improve its segmentation of skin lesions from their surrounding healthy skin. In this manner, the non-star shape segments in FCN prediction
maps are penalized to guarantee a global structure in segmentation results. The experimental results on the ISBI 2017 skin segmentation
challenge dataset demonstrate the advantage of regularizing FCN parameters by
the star shape prior.



More examples in this category can be found in gland segmentation \cite{bentaieb2016topology}, kidney segmentation \cite{ravishankar2017learning}, liver segmentation \cite{zheng2019semi-supervised} and cardiac segmentation \cite{oktay2017anatomically,zotti2019convolutional}.

\textbf{Learning anatomical priors via generative models}

In the third approach, the anatomical priors (especially the shape prior) are learned by some generative models first and then incorporated into segmentation networks. 

For example, in the cardiac MR segmentation process, a shape multi-view autoencoder (MAE) is proposed to learn shape priors from MR images of multiple standard views \cite{chen2019learning}. The information encoded in the latent space of the trained shape MAE is incorporated into multi-view U-Net (MV U-Net) in the fuse block to guide the segmentation process. 


Another example is shown in \cite{luo2020shape}, where the shape constrained network (SCN) is proposed to incorporate the shape prior into the eye segmentation network. More specifically, the prior information is first learned by a VAE-GAN, and then the pre-trained encoder and discriminator are leveraged to regularize the training process.

More examples can be found in brain geometry segmentation in MRI \cite{dalca2018anatomical}, 3D fine renal artery segmentation \cite{he2019dpa-densebiasnet},
overlapping cervical cytoplasms segmentation \cite{song2020shape}, scapula segmentation \cite{boutillon2020combining}, liver segmentation \cite{pham2020liver}, carotid segmentation \cite{engin2020agan}, and head and neck segmentation \cite{gao2020focusnetv2}.

\subsubsection{Other Hand-crafted Features}

Besides anatomical priors, some hand-crafted features are also utilized for segmentation tasks. Generally speaking, there are two ways to incorporate the hand-crafted features into deep learning models: the feature-level fusion and the input-level fusion.

In the feature-level fusion, the hand-crafted features and the features learned by the deep models are concatenated.  For example, for the gland segmentation in histopathology images \cite{rezaei2019gland}, two handcrafted features, namely invariant LBP features as well as $H\&E$ components, are firstly calculated from images. Then these features are concatenated with the features generated from the last convolutional layer of the network for predicting the segmentation results. Similarly, in the brain structure segmentation \cite{kushibar2018automated}, the spatial atlas prior is first represented as a vector and then concatenated with the deep features.

For the input-level fusion, the hand-crafted features are transformed into the input patches. Then the original image patches and the feature-transformed patches are fed into a deep segmentation network. For example in \cite{khan2020cascading}, for automatic brain tumor segmentation in MRI images, three handcrafted features (i.e., mean intensity, LBP and HOG) are firstly extracted. Based on these features, a SVM is employed to generate confidence surface modality (CSM) patches. Then the CSM patches and the original patches from MRI images are fed into a segmentation network. This method achieves good performance on BRATS2015 dataset.

\begin{table*}[]
\small
\center
\caption{The comparison of the quantitative metrics for some medical object segmentation methods after incorporating domain knowledge}
\begin{tabular}{|l|l|l|}
\cline{1-3}
\multicolumn{1}{|c|}{Reference}                              & \multicolumn{1}{c|}{\begin{tabular}[c]{@{}l@{}}Baseline Model/With Domain Knowledge\end{tabular}}                              & Dice score             \\ \hline
\multirow{1}{*}{\cite{chen2019learning}} & \multicolumn{1}{c|}{3D U-Net/MV U-Net} & 0.923/0.926     \\ \hline
\multirow{1}{*}{\cite{he2019dpa-densebiasnet}} & \multicolumn{1}{c|}{V-Net/DPA-DenseBiasNet} & 0.787/0.861     \\ \hline
\multirow{1}{*}{\cite{jiang2018tumor}} & \multicolumn{1}{c|}{\begin{tabular}[c]{@{}l@{}}Masked cycle-GAN/Tumore\\ aware semi-unsupervised\end{tabular}} & 0.630/0.800    \\ \hline
\multirow{1}{*}{\cite{valindria2018multi}} & \multicolumn{1}{c|}{\begin{tabular}[c]{@{}l@{}}modality specific method/dual-stream \\ encoder-decoder multi-model method\end{tabular}} & 0.838/0.860     \\ \hline
\multirow{1}{*}{\cite{yue2019cardiac}} & \multicolumn{1}{c|}{U-Net/SRSCN} & 0.737/0.830    \\ \hline
\multirow{1}{*}{\cite{yu2019annotation}} & \multicolumn{1}{c|}{\begin{tabular}[c]{@{}l@{}}U-Net/SC-GAN\end{tabular} } & 0.742/0.824     \\ \hline
\multirow{1}{*}{\cite{zhang2018translating}} & \multicolumn{1}{c|}{\begin{tabular}[c]{@{}l@{}} cycle- and shape-consistency GAN\\ trained without/with synthetic data\end{tabular}} & 0.678/0.744     \\ \hline

\end{tabular}
\label{tab:seg_quan}
\end{table*}

In addition, using handcrafted features by input-level fusion is also adopted in cell nuclei segmentation \cite{narotamo2020combining}. In particular, as nuclei are expected to have an approximately round shape, a map of gradient convergence is computed and be used by CNN as an extra channel besides the fluorescence microscopy image. Experimental results show higher F1-score when compared with other methods. Another example in this category can be found in brain tumor segmentation \cite{khan2020cascading}.

\subsection{Summary}
\label{sec:seg_overview}
The aforementioned sections described researches of incorporating domain knowledge for object (lesion or organ) segmentation in medical images. 
The segmentation performance of some methods is shown in Table \ref{tab:seg_quan}, where the Dice score is used with a higher score indicating a better performance.

We can see that similar to disease diagnosis, using information from natural images like ImageNet is quite popular for lesion and organ segmentation tasks. The reason behind it may be that segmentation can be seen as a specific classification problem. Meanwhile, besides the ImageNet, some video datasets can also be utilized for segmenting 3D medical images (e.g.,  \cite{zeng20173d}). Using extra medical datasets with different modalities has also been proven to be helpful, although most applications are limited in using MRI to help segmentation tasks in CT images \cite{valindria2018multi}. Leveraging domain knowledge from medical doctors is also widely used in segmentation tasks. In particular, the anatomical priors of organs are widely adopted. However, anatomical priors are only suitable for the segmentation of organs with fixed shapes like hearts \cite{chen2019learning} or lungs \cite{chen2019learning}.

\section{Other Medical Applications}
\label{sec:otherappl}

In this section, we briefly introduce the works on incorporating domain knowledge in other medical images analysis applications, like medical image reconstruction, medical image retrieval and medical report generation.
\subsection{Medical Image Reconstruction}
\label{sec:otherappl_reconstruction}

The objective of medical image reconstruction is reconstructing a diagnostic image from a number of measurements (e.g., X-ray projections in CT or the spatial frequency information in MRI). Deep learning based methods have been widely applied in this field \cite{qin2018convolutional,schlemper2017deep}. It is also quite common that external information is incorporated into deep learning models for medical image reconstruction.



Some methods incorporate hand-crafted features in the medical image reconstruction process. For example, a network model called as DAGAN is proposed for the reconstruction of compressed sensing magnetic resonance imaging (CS-MRI) \cite{yang2017dagan}. In the DAGAN, to better preserve texture and edges in the reconstruction process, the adversarial loss is coupled with a content loss. In addition, the frequency-domain information is incorporated to enforce similarity in both the image and frequency domains. Experimental results show that the DAGAN method provides superior reconstruction with preserved perceptual image details.

In \cite{yedder2019limited}, a new image
reconstruction method is proposed to solve the limited-angle and limited sources breast cancer diffuse optical tomography (DOT) image reconstruction
problem in a strong scattering medium.  By adaptively focusing on important features and filtering irrelevant and
noisy ones using the Fuzzy Jaccard loss, the network is able to reduce false
positive reconstructed pixels and reconstruct more accurate images.


Similarly, a GAN-based method is proposed to recover MRI images of the target contrast \cite{dar2018synergistic}. The method simultaneously leverages the relatively low-spatial-frequency information available in the collected evidence for the target contrast and the relatively high-spatial frequency information available in the source contrast. Demonstrations on brain MRI datasets indicate the proposed method outperforms state-of-the-art reconstruction methods, with enhanced recovery of high-frequency tissue structure, and improved reliability against feature leakage or loss.




\subsection{Medical Image Retrieval}
\label{sec:otherappl_retrieval}

The hospitals are producing large amount of imaging data and the development of medical image retrieval, especially the content based image retrieval (CBIR) systems can be of great help to aid clinicians in browsing these large datasets. 
Deep learning methods have been applied to CBIR and have achieved high performance due to their superior capability for extracting features automatically.

It is also quite common that these deep learning models for CBIR utilize external information beyond the given medical datasets. Some methods adopt the transfer learning to utilize the knowledge from natural images or external medical datasets \cite{ahmad2017sinc,khatami2018sequential,swati2019content}. For example, the VGG model pre-trained based on ImageNet is used in brain tumor retrieval process \cite{swati2019content}, where a block-wise fine-tuning strategy is proposed to enhance the retrieval performance on the T1-weighted CE-MRI dataset. Another example 
can be found in x-ray image retrieval process \cite{khatami2018sequential}, where a model pre-trained on the large augmented dataset is fine-tuned on the target dataset to extract general features. 

Besides, as features play an important role in the similarly analysis in CBIR, some methods fuse prior features with deep features. In particular, in the chest radiograph image retrieval process, the decision values of binary features and texture features are combined with the deep features in the form of decision-level fusion \cite{anavi2015comparative}. Similarly, the metadata such as patients' age and gender is combined with the image-based features extracted from deep CNN for X-ray chest pathology image retrieval \cite{anavi2016visualizing}. Furthermore, the features extracted from saliency areas can also be injected into the features extracted from the whole image for the high retrieval accuracy \cite{ahmad2017sinc}.

\subsection{Medical Report Generation}
\label{sec:otherappl_reports}
Recently, deep learning models for image captioning have been successfully applied for automatic generation of medical reports \cite{jing2017automatic,liu2019clinically}. It is also found that incorporating external knowledge can help deep learning models to generate better medical reports. 


For example, some methods try to incorporate specific or general patterns that doctors adopt  when writing reports. For example, radiologists generally write reports using certain templates. Therefore, some templates are used during the sentence generation process \cite{li2018hybrid,li2019knowledge}. Furthermore, as the explanation given by doctors is fairly simple and phrase changing does not change their meaning, a model-agnostic method is presented to learn the short text description to explain this decision process \cite{gale2018producing}.

In addition, radiologists follow some procedures when writing reports: they generally first check a patient's images for abnormal findings, then write reports by following certain templates, and adjust statements in the templates for each individual case when necessary \cite{hong2013content}. This process is mimicked in \cite{li2019knowledge}, which first transfers the visual features of medical images into an abnormality graph, then retrieves text templates based on the abnormalities and their attributes for chest X-ray images.

In \cite{zhang2020radiology}, a pre-constructed graph embedding module (modeled with a graph CNN) on multiple disease findings is utilized to assist the generation of reports. The incorporation of knowledge graph allows for dedicated feature learning
for each disease finding and the relationship modeling between them. Experiments on the publicly accessible dataset (IU-RR) demonstrate the superior performance of the method integrated with the proposed graph module.

\section{Research Challenges and Future Directions}
\label{sec:future_work}

The aforementioned sections reviewed research studies on deep learning models that incorporate medical domain knowledge for various tasks. Although using medical domain knowledge in deep learning models is quite popular, there are still many difficulties about the selection, representation and incorporating method of medical domain knowledge. In the following sections, we summarize challenges and future directions in this area.


\subsection{The Challenges Related to the Identification and Selection of Medical Domain Knowledge}

Identifying medical domain knowledge is not an easy task. Firstly, the experiences of medical doctors are generally subjective and fuzzy. Not all medical doctors can give accurate and objective descriptions on what kinds of experiences they have leveraged to finish a given task. In addition, experiences of medical doctors can vary significantly or even contradictory to each other.  Furthermore, medical doctors generally utilize many types of domain knowledge simultaneously. 
Finally, currently the medical domain knowledge is identified manually, and there is no existing work on the automatically and comprehensively identifying medical domain knowledge for a given area.


One solution to the automatic identifying medical knowledge is through text mining techniques on the guidelines, books, and medical reports related to different medical areas. Guidelines or books are more robust than individual experiences. Medical reports generally contain specific terms (usually adjectives) that describe the characteristics of tumors.  These terms, containing important information to help doctors to make diagnosis, can potentially be beneficial for deep learning models.

Besides the identification of medical domain knowledge, how to select appropriate knowledge to help image analysis tasks is also challenging. It should be noted that a common practice of medical doctors may not be able to help deep learning models because \emph{the domain knowledge might be learned by the deep learning model from training data.} We believe that the knowledge that is not easily learned by a deep learning model can greatly help the model to improve its performance.

\subsection{The Challenges Related to the Representation of Medical Domain Knowledge}

The original domain knowledge of medical doctors is generally in the form of descriptive sentences like `we will focus more on the margin areas of a tumor to determine whether it is benign or malignant', or `we often compare bilateral images to make decision'. How to transform the knowledge into appropriate representations and incorporate it into deep learning models need a careful design.


There are generally four ways to represent a certain type of medical domain knowledge. One is to represent knowledge as patches or highlighted images (as in \cite{tan2019expert}). This is generally used when doctors pay more attention to specific areas. The second approach is to represent knowledge as feature vectors \cite{majtner2016combining}. This way is suitable when the selected domain knowledge can be described as certain features.
The third approach is to represent domain knowledge as extra labels \cite{hussein2017risk,liu2018integrate}, which is suitable for the knowledge in clinical reports or extra feature attributes of diseases. The last approach is to embed medical domain knowledge in network structure design, which is suitable to represent high-level domain knowledge like the training pattern and diagnostic pattern of medical doctors\cite{jimenez2019medical,liu2019from,2020Cross}.

\subsection{The Challenges Related to the Incorporating Methods of Medical Domain Knowledge}

Currently, there are four ways to incorporate medical domain knowledge.  The first is to transform the knowledge into certain patches or highlighted images and put them as extra inputs \cite{xie2019knowledge-based}. The second approach is via concatenation \cite{rezaei2019gland}. The domain knowledge are generally transformed into feature vectors and concatenated with those extracted by deep learning models. The third way is the attention mechanism \cite{li2019attention}. The approach is applicable when doctors focus on certain areas of medical images or focus on certain time slots on medical videos. The last one is to learn the domain knowledge by using some specific network structures like generative models \cite{chen2019learning,luo2020shape}.




However, most of the existing works only incorporate a single type of medical domain knowledge, or a few types of medical domain knowledge of the same modality (e.g., a number of hand-crafted features). In practice, experienced medical doctors usually combine different experience in different stages.

There are some researches that simultaneously introduce high-level domain knowledge (e.g.,  diagnostic pattern, training pattern) and the low-level one (e.g.,  hand-crafted features, anatomical priors). In particular, the high-level domain knowledge is incorporated as input images, and low-level one is learned by using specific network structures \cite{chen2019learning}. In addition, besides incorporating into network directly, the information from low-level domain knowledge can also be used to design the training orders when combined with the easy-to-hard training pattern \cite{tang2018attention}. We believe that  simultaneously incorporating multiple kinds of medical domain knowledge can better help deep learning models in various medical applications. 

\subsection{Future Research Directions}
\label{sec:seg_disease}

Besides the above challenges, there are several directions that we feel need further investigation in the future.

\textbf{Domain adaptation}

Domain adaptation is developed to transfer the information from a source domain to a target one. Via techniques like adversarial learning \cite{goodfellow2014generative}, domain adaptation is able to narrow the domain shift between the source domain and the target one in input space \cite{hoffman2017cycada}, feature space \cite{long2016unsupervised,tzeng2017adversarial} and output space \cite{luo2019taking,tsai2018learning}. 
It can be naturally adopted to transfer knowledge of one medical dataset to another \cite{li2020dual}, even when they have different imaging modes or belong to different diseases \cite{ghafoorian2017transfer,jiang2018tumor}.

In addition, unsupervised domain adaptation (UDA) is a promising avenue to enhance the performance of deep neural networks on the target domain, using labels only from the  source domain. This is especially useful for medical field, as annotating the medical images is quite labor-intensive and the lack of annotations is quite common in medical datasets.  Some examples have demonstrated the effectiveness of UDA in disease diagnosis and organ segmentation \cite{chen2018semantic,yang2019dscgans,zhang2020collaborative,liu2020pdam}, but further depth study needs to be implemented in the future.

\textbf{The knowledge graph}

We believe that the knowledge graph \cite{wang2014knowledge}, with the character of embedding different types of knowledge, is a generic and flexible approach to incorporate multi-modal medical domain knowledge. Although rarely used at present, it also shows advantage in medical image analysis tasks, especially in medical report generation \cite{li2019knowledge}. We believe that more types of knowledge graph can be used to represent and learn domain knowledge in medical image analysis tasks.

According to different relationships in graphs, there are three possible types of knowledge graphs can be established. The first knowledge graph reflects the relationship among different kinds of medical domain knowledge with respect to a certain disease. This knowledge graph can help us identify a few key types of knowledge that may help to improve the performance deep learning models. The second type of knowledge graph may reflects the relationship among different diseases. This knowledge graph can help us to find out the potential domain knowledge from other related diseases. The third type one can describe  the relationship among medical datasets. These datasets can belong to different diseases and in different imaging modes (e.g., CT, MRI, ultrasound). This type of knowledge graph will help to identify the external datasets that may help to improve the performance of the current deep learning model.

\textbf{The generative models}

The generative models, like GAN and AE, have shown great promise to be applied to incorporate medical domain knowledge into deep learning models, especially for segmentation tasks.  GAN has shown its capability to leverage information from extra datasets with different imaging modes (e.g., using a MRI dataset to help segmenting CT images \cite{chen2018semantic,jiang2018tumor}). In addition, GAN is able to learn important features contained in medical images in a weakly or fully unsupervised manner and therefore is quite suitable for medical image analysis.

AE-based models have already achieved a great success in extracting features, especially the shape priors in objects like organs or lesions in a fully unsupervised manner \cite{chen2019learning,luo2020shape}. The features learning by AE can also be easily integrated into the training process of networks.

\textbf{Network architecture search (NAS)}

At last, we mentioned in the previous section that one challenge is to find appropriate network architectures to incorporate medical domain knowledge. We believe one approach to address this problem is the technique of network architecture search (NAS).  NAS has demonstrated its capability to automatically find a good network architecture in many computer vision tasks \cite{wistuba2019a} and has a great promise in the medical domain \cite{guo2020organ}. For instance, when some hand-crafted features are used as the domain knowledge, with the help of NAS, a network structure can be identified with the special connections between domain knowledge features and deep features. In addition, instead of designing the feature fusion method (feature-level fusion, decision-level fusion or input-level fusion) for these two kinds of features, the integrating phase and integrating intensity of these two kinds of features can also be determined during the searching process.
\section{Conclusion}
\label{sec:conclusion}

In this paper, we give a comprehensive survey on incorporating medical domain knowledge into deep learning models for various medical image analysis tasks ranging from disease diagnosis, lesion, organ and abnormality detection to lesion and organ segmentation. In addition, some other tasks such as medical image reconstruction, medical image retrieval and medical report generation are also included. For each task, we first introduce different types of medical domain knowledge, and then review some works of introducing domain knowledge into target tasks by using different incorporating methods. From this survey, we can see that with appropriate integrating methods, different kinds of domain knowledge can help deep learning models to better accomplish corresponding tasks.

Besides reviewing current works on incorporating domain knowledge into deep learning models, we also summarize challenges of using medical domain knowledge, and introduce the identification, selection, representation and incorporating method of medical domain knowledge. Finally, we give some future directions of incorporating domain knowledge for medical image analysis tasks.

\section*{Acknowledgments}
This work was supported by the National Natural Science Foundation of China [grant numbers 61976012, 61772060]; the National Key R\&D Program of China [grant number 2017YFB1301100]; and the CERNET Innovation Project [grant number NGII20170315].

\bibliographystyle{IEEEtran}
\bibliography{domain_knowledge}
%



\end{CJK}
\end{document}